\documentclass[pre,aps,floatfix,showpacs,preprint]{revtex4-1}

\usepackage{graphicx}
\usepackage{amsmath}
\usepackage{float}
\begin{document}

\title{Influence of Rough and Smooth Walls on Macroscale Flows in Tumblers}

\author{Umberto D'Ortona}
\email{umberto@l3m.univ-mrs.fr}
\author{Nathalie Thomas} 
\affiliation{Aix-Marseille Universit\'e, CNRS, Centrale Marseille, M2P2 UMR 7340, 13451, Marseille, France}
\author{Zafir Zaman}
\affiliation{Department of Chemical and Biological Engineering,
         Northwestern University, Evanston, Illinois 60208, USA}
\author{Richard M. Lueptow}
\affiliation{Department of Mechanical Engineering, 
         Northwestern University, Evanston, Illinois 60208, USA}
\affiliation{The Northwestern Institute on Complex Systems (NICO), Northwestern University, Evanston, Illinois 60208, USA}

\date{\today}
\begin{abstract}
Walls in discrete element method (DEM) simulations of granular flows are 
sometimes modeled as a closely packed monolayer of fixed particles, resulting in
a rough wall rather than a geometrically smooth wall.  An implicit
assumption is that the resulting rough wall differs from a smooth wall only
locally at the particle scale.  Here we test this assumption by considering
the impact of the wall roughness at the periphery of the flowing layer on
the flow of monodisperse particles in a rotating spherical tumbler. We find
that varying the wall roughness significantly alters average particle
trajectories even far from the wall.  Rough walls induce greater poleward
axial drift of particles near the flowing layer surface, but decrease the
curvature of the trajectories. Increasing the volume fill level in the
tumbler has little effect on the axial drift for rough walls, but increases
the drift while reducing curvature of the particle trajectories for smooth 
walls. The mechanism for these effects is related to the degree of local 
slip at the bounding wall, which alters the flowing layer thickness near
the walls, affecting the particle
trajectories even far from the walls near the equator of the tumbler.  Thus,
the proper choice of wall conditions is important in the accurate
simulation of granular flows, even far from the bounding wall.
\end{abstract}

\maketitle
\section{Introduction}

Discrete element method (DEM) simulations have been used extensively to study
the motion of granular materials in many situations as a predictive tool as
well as to obtain data that is otherwise inaccessible experimentally. In the
method each particle's motion is governed by Newton's laws: the goal is to
compute the evolution of linear and angular momentum of every individual
particle by using appropriate contact force models 
\cite{SchaferDippel96,Ristow00,CundallStrack79}. 
While early DEM simulations could manage systems
with only a few hundred to tens of thousands of particles \cite{Shoichi98,HirshfeldRapaport01,Ristow94,McCarthyOttino98,MoakherShinbrot00,YangZou03,YamaneNakagawa98},
simulating millions of particles is now practical with
advances in computer technology.

As with many simulation approaches, one of the key aspects is the
implementation of boundary conditions.  Two types of wall boundary conditions
can be implemented in DEM simulations for the calculation of the
collision force between mobile granular particles and the walls:
1) geometrically smooth surfaces, which 
are assumed to have infinite mass and a specified radius of curvature 
(infinite for planar walls) (for example, see \cite{ChenLueptow11,MoakherShinbrot00,TaberletLosert04,Rapaport02,TaberletNewey06}); 
and 2) a geometrically
rough surface made up of a closely packed monolayer of fixed particles
conforming to the geometry of the wall surface (for example, see 
\cite{DaCruzEmam05,McCarthyOttino98,PoschelBuchholtz95,JuarezChen11,BertrandLeclaire05}). The latter approach, often called a “rough wall,” 
is 
easy to implement because interactions between mobile particles and immobile
 wall particles are modeled in almost exactly the same way as between pairs of
mobile particles.  The only difference is that the wall particles remain
fixed in their wall position.  An implicit assumption is typically that a
 rough wall differs from a smooth wall only locally at the particle scale
or perhaps through the thickness of the flowing layer, which is typically 
$O(10)$ particles thick \cite{PignatelLueptow12}, but is unlikely to have a 
global effect on the flow, particularly far from the wall.  Likewise, it is
 often implicitly assumed that a rough wall is
similar to a smooth wall, but one with a very high coefficient of friction.

\begin{figure}[htbp]
\includegraphics[width=0.7\linewidth]{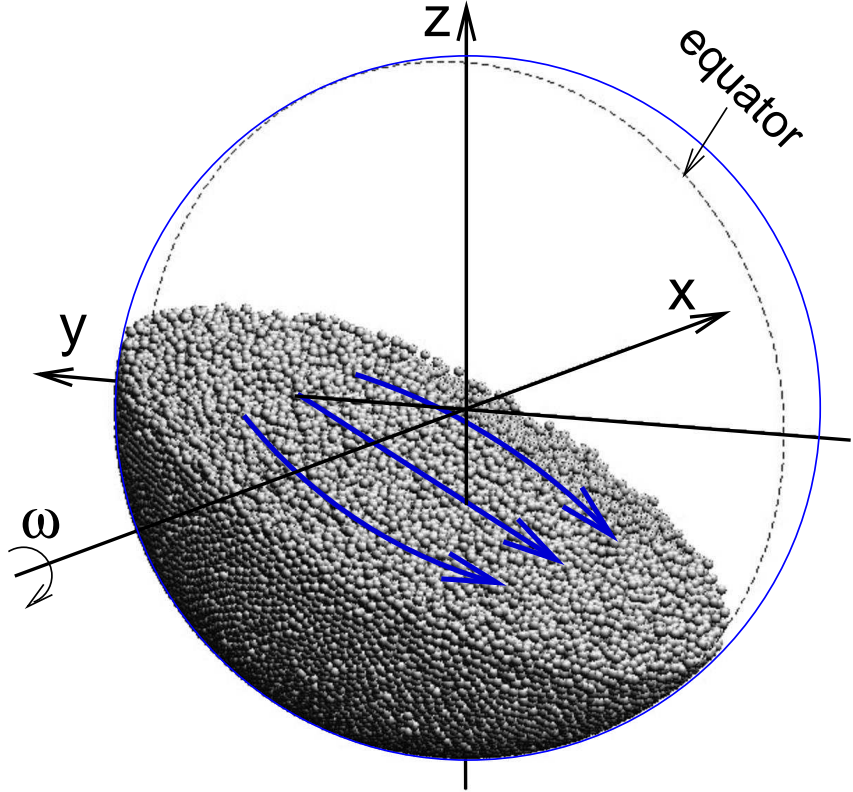}
\caption{(Color online) 14~cm diameter spherical tumbler filled at 30\% with 2~mm particles.
The blue arrows at the free surface show the direction of the flow. $x$ is the polar axis, $y$ and $z$ are in the equatorial plane.}
\label{spherecoord}
\end{figure}

In this paper, we examine the impact of the wall boundary condition on the flow
using the system of monodisperse
particles in a partially-filled spherical tumbler rotating with angular
velocity $\omega$ about a horizontal axis (Fig.~\ref{spherecoord}).
We consider the situation where the free surface is essentially
flat and continuously flowing. In this regime, the surface of the flowing
layer maintains a dynamic angle of repose $\beta$  with respect to horizontal,
which depends on the frictional properties and diameter $d$ of the particles
and the rotational speed of the tumbler 
\cite{TaberletRichard03,duPontGondret03,PignatelLueptow12,MeierLueptow07}. 

In spherical tumblers, it has been shown that monodisperse particles slowly 
drift axially toward the pole near the surface of the flowing layer, while
they drift axially toward the equator deep in the flowing layer with an axis
of symmetry at the equator \cite{ZamanDOrtona13}.
\begin{figure}[htbp]
\includegraphics[width=0.7\linewidth]{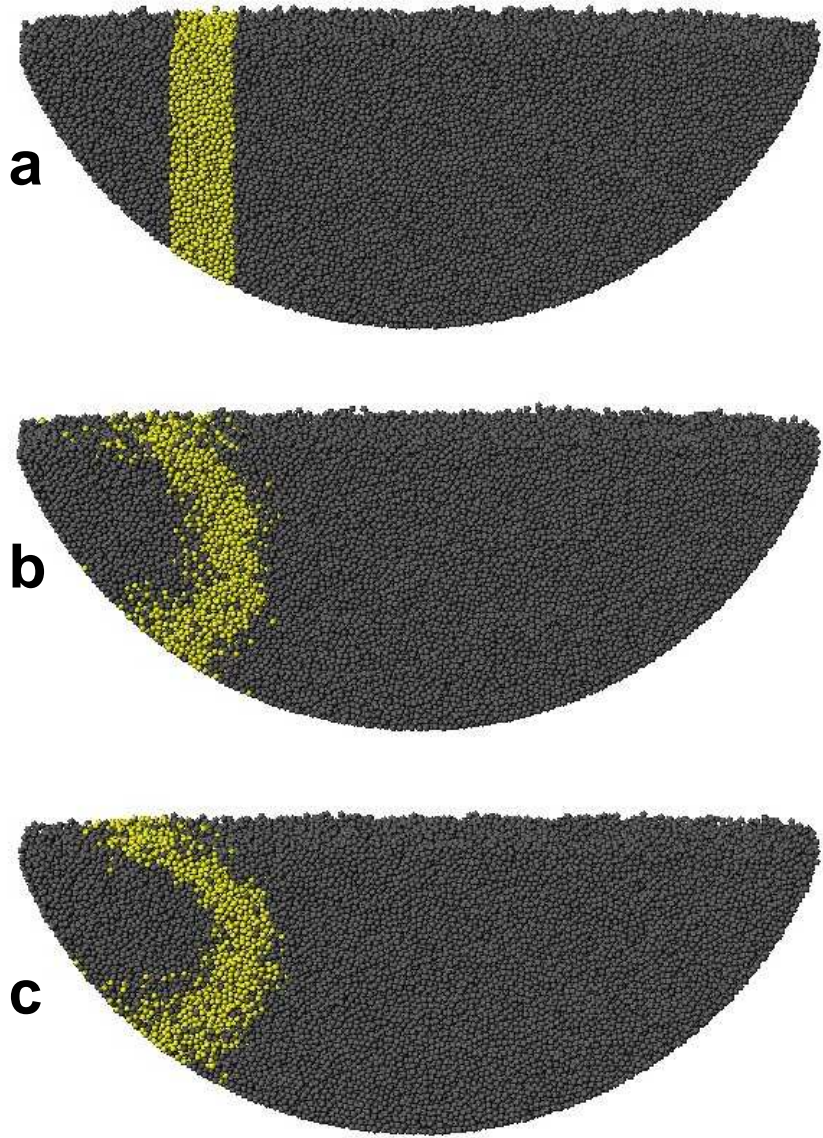}
\caption{(Color online) Axial deformation of a vertical band of 1~mm colored particles in a 
14~cm diameter spherical tumbler filled at 30\% and rotating at 15~rpm. 
(a) The band is initially placed between 
-4~cm and -3~cm from the equator. After rotation of the tumbler about a 
horizontal axis 
in the plane of the page so that particles have made two circulations through
the flowing layer the band
is deformed. The wall is (b) perfectly smooth or (c) rough, made of 1~mm 
particles. The front half of the bed of particles has been removed in order 
to view a cross-section through the bed of particles.}
\label{drift1mm}
\end{figure}
For example, consider the deformation of an initially vertical 
band of colored particles in a 30\% full spherical tumbler 
shown in Fig.~\ref{drift1mm}(a). In this
figure, the axis of rotation is horizontal, and the front half of the bed
of 1~mm diameter particles has been removed to view a cross-section of the particles. 
In Fig.~\ref{drift1mm}, after 2 circulations of particles through the flowing 
layer, 
the initially vertical band of colored particles is more slightly deformed in 
the case of a rough wall 
(1~mm diameter particles forming the wall) in Fig.~\ref{drift1mm}(c) than
in the case of a smooth wall in Fig.~\ref{drift1mm}(b). To generate these 
images, it was necessary to choose
a time corresponding to an integer number of circulations of particles through
the flowing layer so that particles that were in the static bed return to
the static bed and particles in flowing layer return to the flowing layer.
To do this, we 
measured the average recirculation time of all particles: 
2.14~s for the smooth case and 2.04~s for the rough case. 
This recirculation time difference does not come from slip between 
the smooth wall and the granular media in solid body rotation.
This difference is instead due to a global modification of the flow
due to the wall roughness.

The deformation of the colored band comes about from asymmetries in the curvature of 
particle trajectories in the flowing layer that result in poleward drift near 
the free surface \cite{ZamanDOrtona13}.  Mass conservation requires 
equator-directed 
drift deeper in the flowing layer. Although our previous study 
 \cite{ZamanDOrtona13} indicated that the axial drift occurs for both 
smooth and rough walls, the degree of axial drift was not studied in detail.
It is the unexpectedly strong influence of wall roughness on 
deformation of a labeled band in Fig.~\ref{drift1mm}, that we examine 
primarily through
simulations in order to better understand
the impact of wall boundary conditions in DEM simulations and experiments.

By way of background, the wall roughness has been shown to affect the flow
of granular media in cases like chute flows and shear cell flows. 
For instance, 
the roughness of the bottom wall in chute flow has
an impact on velocity \cite{GoujonThomas03,Augenstein78,ThorntonWeinhart12,WeinhartThornton12} and 
velocity profile \cite{Ancey01}.
In planar shear flows 
\cite{Campbell93a,Campbell93b}, a smooth wall results in a slip velocity at the wall and 
large velocity gradients near the wall, whereas a rough wall results in
  no slip velocity
and a nearly uniform velocity gradients. In both chute flow and planar shear
flow, the impact of wall roughness on the flowing layer in the immediate 
vicinity is quite direct.
The frictional effects at the wall alter the local shear rate at the wall, 
which propagates
through the thickness of the entire flowing layer to affect the velocity 
profile. For granular flows in pipes, wall roughness can prevent
 clogging and jamming regimes by deflecting particles toward the center
of the pipe \cite{VerbuchelnParteli15}.
The case studied here, the wall only contacts the flowing layer at its circular periphery, 
not at the bottom of the flowing layer, which is in contact with the underlying
bed of particles that is rotating in solid body motion with the tumbler. Thus,
the impact of the wall roughness is quite different and much less direct.

\section{DEM Simulations}

For the
DEM simulations, a standard linear-spring and viscous damper force model
\cite{ChenLueptow08,SchaferDippel96,Ristow00,CundallStrack79} was used to calculate
the normal force between two contacting particles:\\ 
\mbox{$F_n^{ij}=[k_n\delta - 2 \gamma_n m_{\rm eff} (V_{ij} \cdot \hat r_{ij})]\hat r_{ij}$}, 
where $\delta$ and $V_{ij}$ are the particle overlap and the relative
velocity $(V_i - V_j)$ of contacting particles $i$ and $j$ respectively; 
$\hat r_{ij}$ is
the unit vector in the direction between particles $i$ and $j$; 
$m_{\rm eff} = m_i m_j/(m_i + m_j)$ is the reduced mass of the two particles; 
$k_n = m_{\rm eff} [( \pi/\Delta t )^2 + \gamma^2_n]$ is the normal stiffness 
and $\gamma_n = \ln e/\Delta t$ is the normal damping, where $\Delta t$
is the collision time and $e$ is the restitution coefficient \cite{ChenLueptow08,Ristow00}. A
standard tangential force model \cite{SchaferDippel96,CundallStrack79} with
elasticity was implemented: $F^t_{ij}= -\min(|\mu F^n_{ij}|,|k_s\zeta|){\rm sgn}(V^s_{ij})$, where
$V^s_{ij}$ is the relative tangential velocity of two particles \cite{Rapaport02},
$k_s$ is the tangential stiffness, $\mu$ the Coulomb friction coefficient and $\zeta(t) = \int^t_{t_0} V^s_{ij} (t') dt'$ is the net
tangential displacement after contact is first established at time $t = t_0$.
The velocity-Verlet algorithm \cite{Ristow00,AllenTildesley02} was used to 
update
the position, orientation, and linear and angular velocities of each
particle. Tumbler walls were modeled as both smooth surfaces (smooth walls)
and as a monolayer of fixed particles that are just touching 
with no overlap between particles (rough walls). 
The
number of particles that comprise the wall ranged from 2300 in the case
of a wall of 6~mm particles to 1.13 million for a wall of 0.25~mm particles. 
All wall conditions
have infinite mass for calculation of the collision force between the
tumbling particles and the wall. 

The spherical tumbler of radius $R=D/2=7$~cm was filled to volume fraction $f$ with 
monodisperse $d=1$~mm to 4~mm particles, though most simulations used $d=2$~mm 
particles; gravitational
acceleration was $g$ = 9.81~m$\,$s$^{-2}$; particle properties correspond to cellulose
acetate: density $\rho =$ 1308 kg m$^{-3}$, restitution coefficient $e$ = 0.87 \cite{DrakeShreve86,FoersterLouge94,SchaferDippel96}. The particles were initially randomly
distributed in the tumbler with a total of about $5\times 10^4$ particles in a
typical simulation. To avoid a close-packed structure, the particles had a
uniform size distribution ranging from 0.95$d$ to 1.05$d$. Unless otherwise 
indicated, the friction
coefficients between particles and between particles and walls was set to 
$\mu = 0.7$.  The collision time was $\Delta t =10^{-4}$ s, consistent with previous
simulations \cite{TaberletNewey06,ChenLueptow11,ZamanDOrtona13} and sufficient for
modeling hard spheres \cite{Ristow00,Campbell02,SilbertGrest07}. These
parameters correspond to a stiffness coefficient $k_n = 7.32\times 10^4$ (N m$^{-1}$)
and a damping coefficient $\gamma_n=0.206 $~kg s$^{-1}$ \cite{SchaferDippel96}.
The integration time step was $\Delta t/50 = 2 \times 10^{-6}$~s to meet the 
requirement of numerical stability \cite{Ristow00}. The rotational
speed of the tumbler is $\omega=15$~rpm in most cases, consistent with 
previous studies of spherical tumbler flow \cite{ChenLueptow09} and chosen such that, for
this system size, the flow is 
continuous, dense, and with a flat free surface rather than discrete 
avalanches at 
very low rotation speeds or a curved free surface at  higher rotation 
speeds. A few cases at other 
rotational speeds were studied ($\omega=2.5$ to 30~rpm), all
in the continuous flow regime. The free 
surface remains flat up to 20~rpm and becomes slightly S-shaped 
at 30~rpm.

\section{Results}

\subsection{Deformation of a vertical band}

\begin{figure}[htbp]
\includegraphics[width=0.85\linewidth]{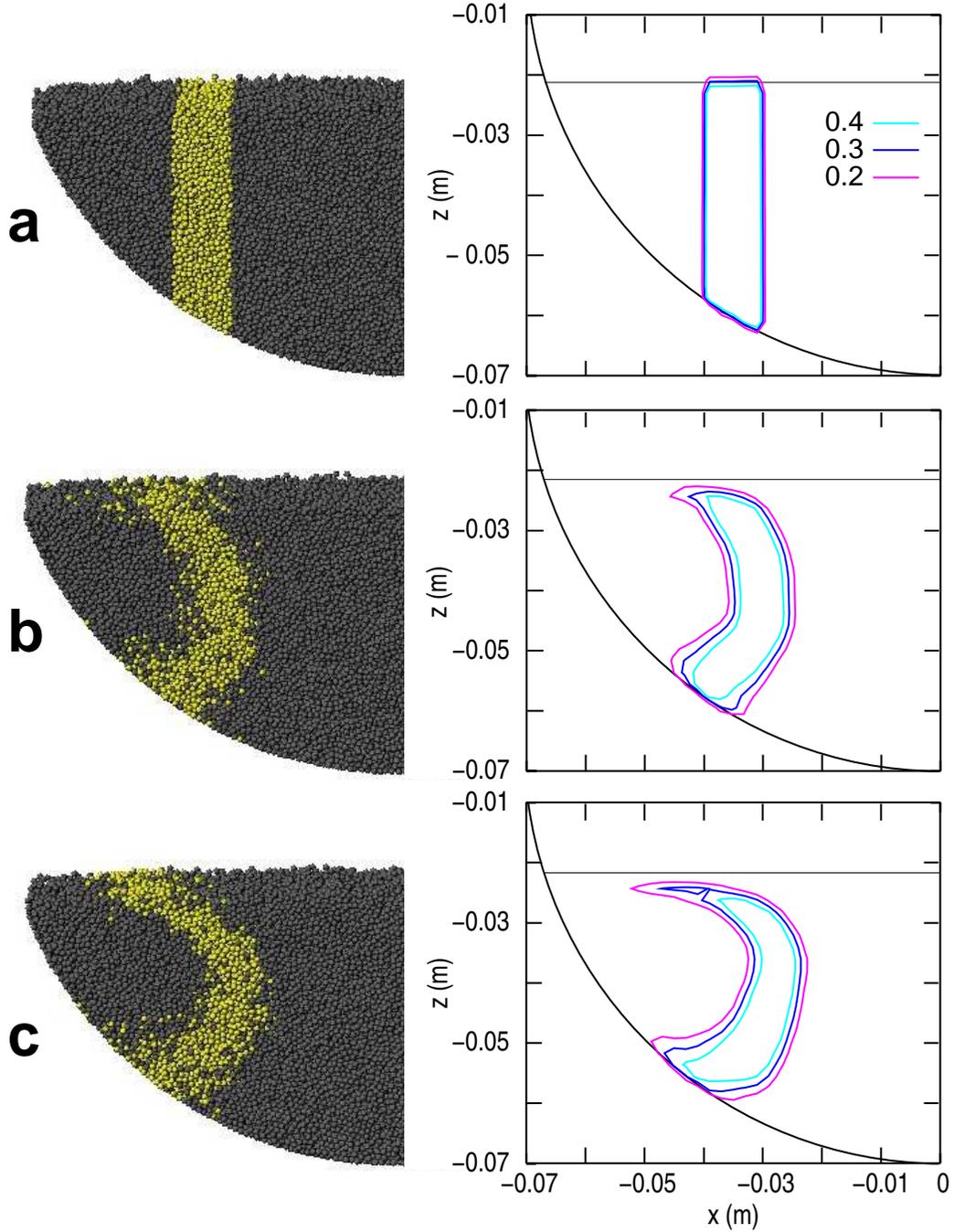}
\caption{(Color online) Deformation of a vertical band of 1~mm colored particles in a 14~cm 
diameter tumbler filled at 30\% and rotating at 15~rpm. The axis of rotation
 is horizontal at $z=0$ 
and in the plane of the page, the equator is at $x=0$, and gravity is down. The image
shows the back half of the particle bed after the front half and right half
are removed. Left: The initial band (a) and
the deformed bands as shown in Fig \ref{drift1mm}. 
Right: the corresponding compacity contour levels of colored particles
to show more accurately the band deformation difference for smooth (b) and
1~mm rough (c) cases. The horizontal line represents the fill level.}
\label{drift1mmcombine}
\end{figure}
Figure~\ref{drift1mmcombine} shows the band deformation, like
Fig.~\ref{drift1mm}, with the corresponding concentration map of the colored 
particles. The contours correspond to iso-volume concentration, or 
iso-compacity. The maximum compacity is about 0.6, but the boundaries 
of the colored particles correspond most closely to contours ranging 
from 0.2 to 0.4. Using this compacity map, the deformation 
of the band is quite clear. 
After just two passes of particles through the flowing layer, the deformation 
of the initially vertical band, with the 1~mm rough wall 
results in more deformation than the band in a tumbler with a smooth wall.  
\begin{figure}[htbp]
\includegraphics[width=0.82\linewidth]{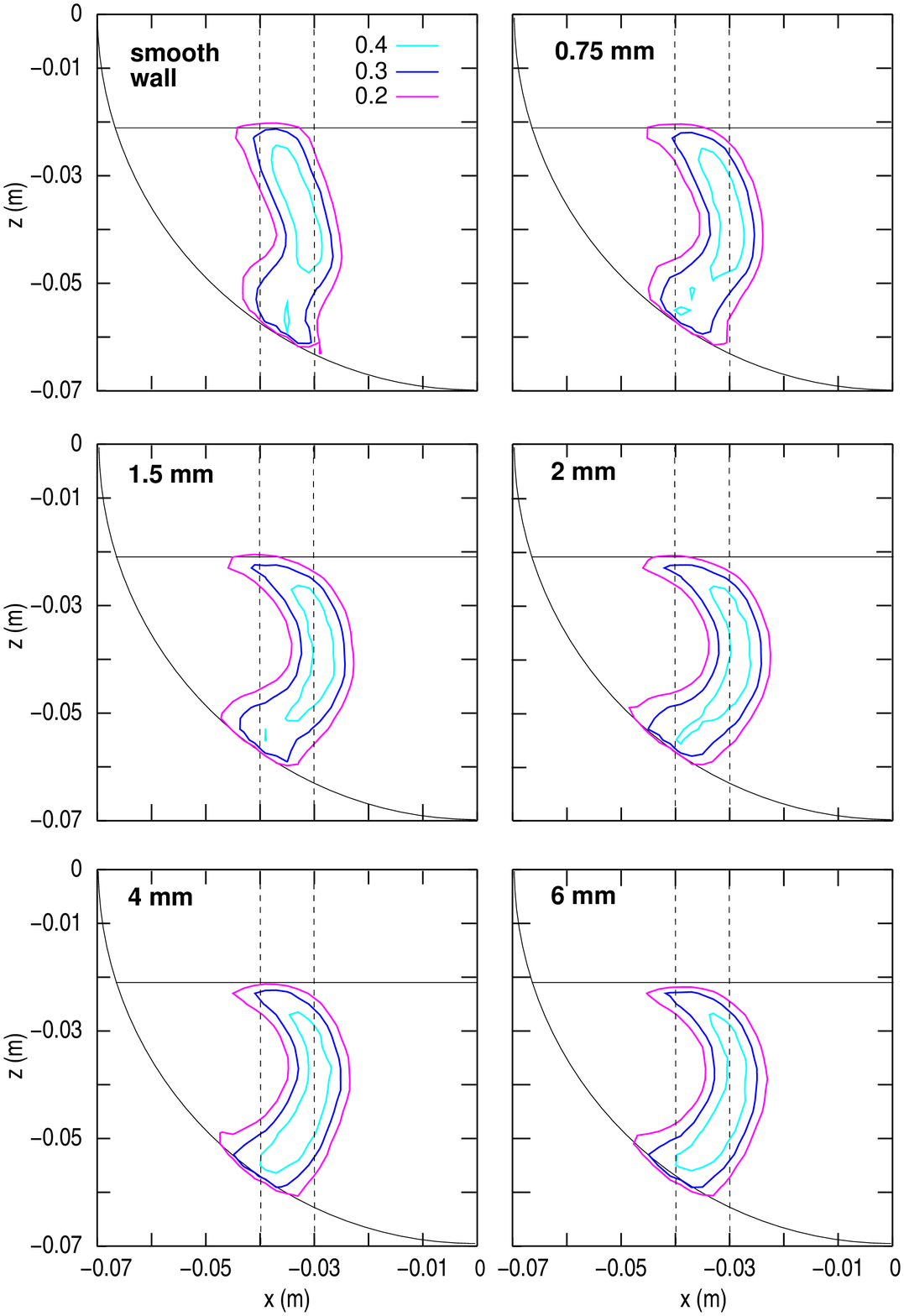}
\caption{(Color online) Deformation of a vertical band of 2~mm colored particles in a 14~cm
diameter tumbler filled at 30\% and rotating at 15~rpm around a horizontal axis
at $z=0$. The band is initially
placed between $x=-4$~cm and $x=-3$~cm from the equator (dashed vertical lines). 
The roughnesses of the 
walls range from smooth to 6~mm particles. The 
iso-compacity contours are measured after two circulations of the particles through 
the flowing layer. The horizontal line represents the fill level.}
\label{shear30}
\end{figure}
Similar results also occur for larger flowing particles. 
Figure~\ref{shear30}  shows the deformation of a band of colored 2~mm particles 
in a 14~cm spherical tumbler filled to 30\% by volume with varying wall roughness so that the innermost surfaces of the wall particles are at a radius $R=7$~cm.
From perfectly smooth to a 2~mm rough wall, the band becomes increasingly
more deformed, though the roughness does 
not modify the band deformation much from 2~mm to 6~mm wall roughness.

\begin{figure}[htbp]
\includegraphics[width=\linewidth]{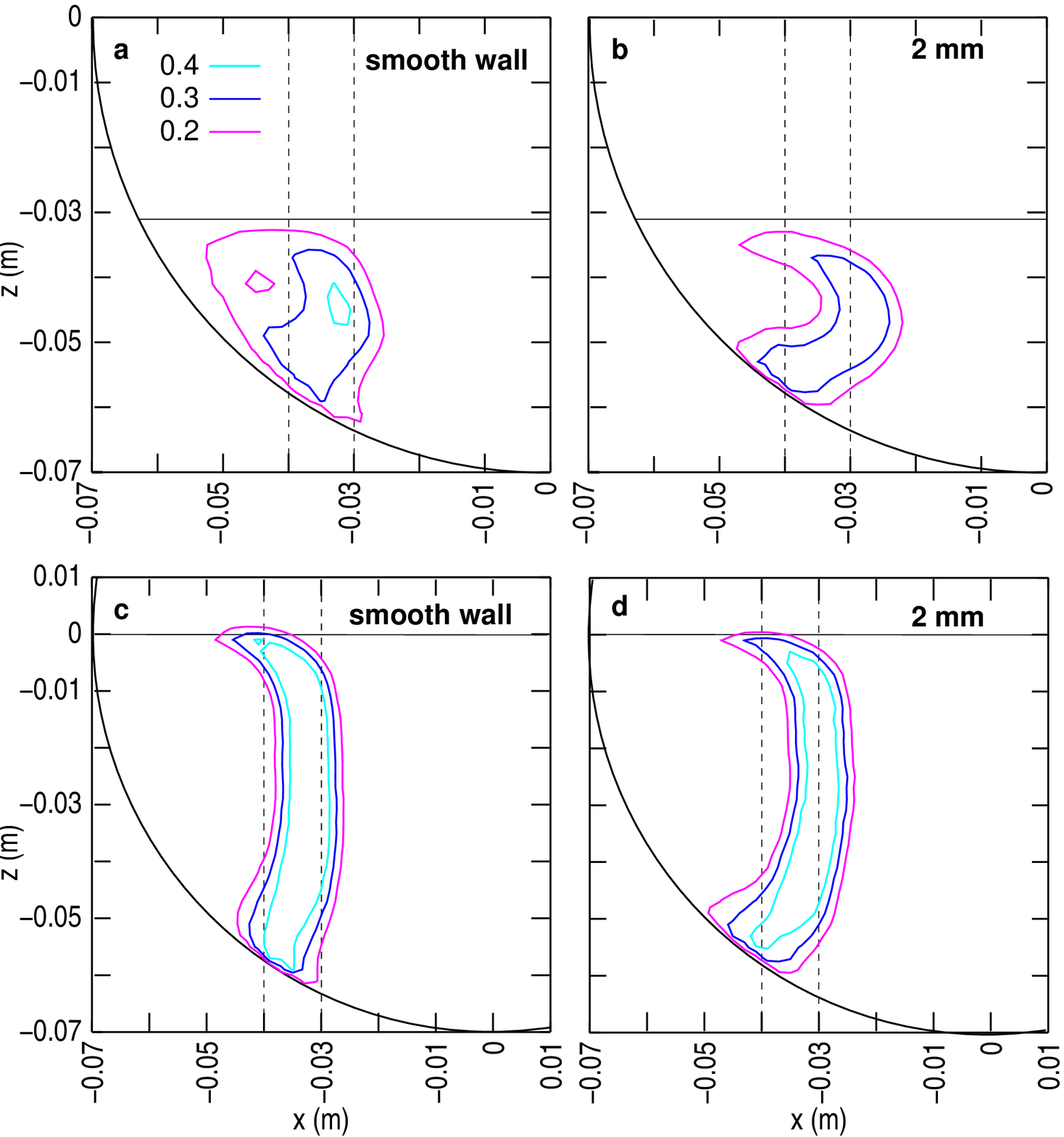}
\caption{(Color online) Deformation of a vertical band of 2~mm colored particles in a 14~cm 
diameter tumbler filled at 20\% (a and b) or 50\% (c and d) and rotating 
at 15~rpm based on iso-compacity contours. The band is initially 
placed between $x=-4$~cm and $x=-3$~cm from the equator (vertical dashed lines). Two 
roughnesses are shown:
smooth wall and 2~mm particles. The horizontal line indicates the fill level.}
\label{shear50}
\end{figure}

Similar results occur for 20\% and 50\% full tumblers as shown in 
Fig.~\ref{shear50}. For each case, only iso-compacity contours for smooth and 
2~mm rough walls have been plotted, 
but we have also simulated wall roughnesses of 0.5, 1, 1.5, and 3~mm with 
consistent results.
In a tumbler filled to 20\%, there is a significant difference in the 
deformation of the colored
band between the smooth and rough cases. For the 50\% full tumbler, the band of
colored particles is less deformed.
In addition, the difference in the band deformation 
between
the smooth and rough walls is smaller for the 50\% fill volume indicating that the influence
of the wall roughness increases for smaller fill fractions.

\subsection{Mean trajectories}

To investigate the mechanism behind the results described above,
we consider 
2~mm particles in a 30\% full 14~cm tumbler with
smooth and rough walls. 
\begin{figure}[htbp]
\includegraphics[width=0.95\linewidth]{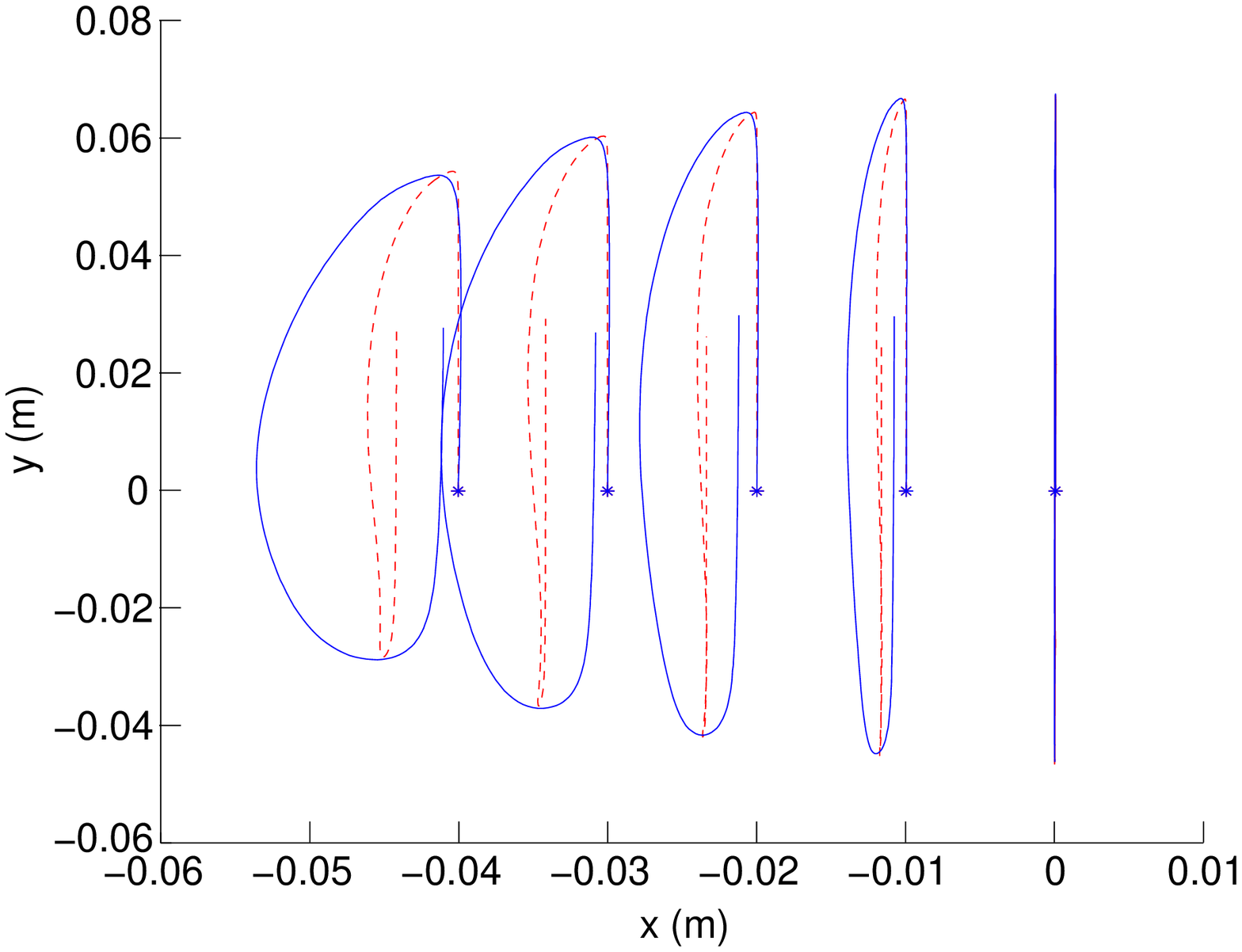}
\caption{(Color online) Comparison (top view) of the mean trajectories 
(initially 2~mm above the sphere wall) for 2~mm particles
starting from the same points in the smooth wall (blue continuous lines) 
and in the rough (2~mm particles) wall (red dashed lines) in a sphere filled 
at 30\%. The horizontal axis of rotation is at $y=0$.}
\label{comparetrajroughsmooth}
\end{figure}
\begin{figure}[htbp]
\includegraphics[width=0.95\linewidth]{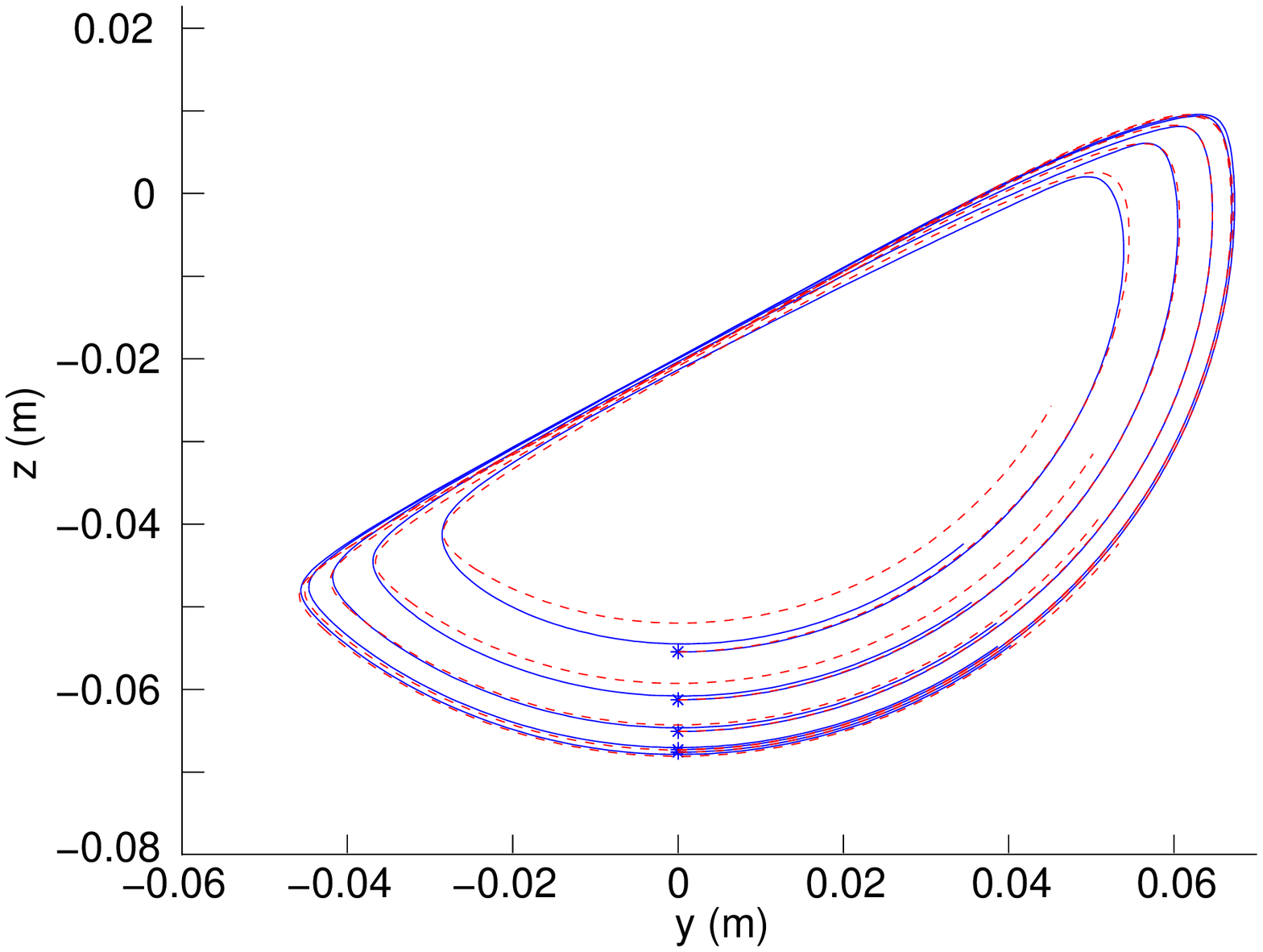}
\caption{(Color online) Comparison (side view) of the mean trajectories 
(initially 2~mm above the sphere wall) for 2~mm particles 
starting from the same points in the smooth wall (blue continuous lines) and 
in the rough (2~mm particles) wall (red dashed lines) in a sphere filled at 
30\%. The axis of rotation is perpendicular to the page at $(y,z)=(0,0)$.}
\label{comparetrajroughsmoothside}
\end{figure}
Figures~\ref{comparetrajroughsmooth} and \ref{comparetrajroughsmoothside}
show average trajectories of particles constructed by integrating the mean
velocity field using a second order Runge-Kutta scheme \cite{ZamanDOrtona13}. 
The two trajectories in each pair correspond
to  smooth and 2~mm rough walls.  The trajectories are shown for several 
different initial axial
positions in the tumbler, where $x=0$ corresponds to the equator.
Both trajectories
in a pair start at the same point (indicated by a star) in the fixed
bed, initially 2~mm away from the sphere wall and below the axis of rotation
(Fig.~\ref{comparetrajroughsmoothside}).
In Fig.~\ref{comparetrajroughsmooth}, the initial vertical portion of the trajectory corresponds to motion in
the fixed bed as the tumbler rotates.  Particles enter the flowing layer
at the topmost part of the trajectory and follow a curved path in the
flowing layer until they re-enter the fixed bed at the bottommost part
of the trajectory, again following a vertical path in the figure when in
solid body rotation.  The paths are viewed looking downward
along the gravity vector, so the flowing layer surface is not perpendicular 
to the line along which the trajectories are viewed.  

Two results are clearly evident in Fig.~\ref{comparetrajroughsmooth}. 
First, the trajectories in the flowing
layer for both cases are curved, with the curvature for smooth walls
much greater than that for rough walls. 
This curvature is
negligible at the equator (at $x=0$, which is a plane of symmetry) 
and increases moving
toward the poles, consistent with previous results in smooth-walled
tumblers \cite{ChenLueptow10,ZamanDOrtona13}.  Second, the trajectories for the smooth-walled tumbler
are nearly closed, displaying only a very small amount of poleward drift
with each pass through the flowing layer.  On the other hand, the
poleward drift for the rough-walled tumbler is much larger. For that
reason, the band undergoes more deformation
in the rough case (Figs.~\ref{drift1mmcombine}-\ref{shear50}). In both
cases, the drift increases toward the poles, consistent with previous
results for smooth tumblers \cite{ZamanDOrtona13}. 
Thus, two quantities can be used 
to characterize the trajectories: the ``displacement'', which is the
maximum axial displacement of the trajectory from its starting point
that occurs at any point during one trajectory circulation through the flowing
layer, and the poleward ``drift'',
which is the net axial displacement of the trajectory after one trajectory
circulation.  Although the displacement and drift differ 
substantially for smooth and 2~mm rough walls, 
Fig.~\ref{comparetrajroughsmoothside} shows there
is little difference in the trajectories viewed along
the axis of rotation. 

While previous simulations and experiments match 
reasonably well when considering axial drift of particles in spherical 
tumblers \cite{ZamanDOrtona13}, we have attempted to 
directly validate the simulation 
results in Fig.~\ref{comparetrajroughsmooth} experimentally using tracer 
particles in a tumbler under similar conditions. However, it is quite difficult 
to obtain quantitative experimental results. Several problems occur. First, visualizing 
the flow and tracking tracer particles in a rough-walled tumbler is 
difficult, because the rough wall makes the particles in the
tumbler challenging to access 
via optical means. Second, the inherent collisional diffusion makes it 
difficult to obtain displacement or drift data because of the statistical 
variation in the tracer particle location, both axially and depthwise. 
Thus, obtaining highly-resolved experimental mean trajectories of tracer 
particles is challenging.  To overcome these problems, we have used an X-ray 
system to track the location of a single X-ray opaque tracer particle 
in a 14~cm spherical tumbler 
that is 30\% filled by volume with 2~mm spherical glass particles rotating at 
6.5~rpm, the maximum speed for which a tracer particle can be accurately 
tracked. The tracer particle is 3~mm, so it 
remains near the surface of the flowing layer, but it still follows the general
trajectory of the 2~mm particles based on direct visual observation in the 
smooth tumbler, an approach we have used successfully in the past 
\cite{ZamanDOrtona13}. The tracer consists of a small ball 
of lead solder enclosed in a plastic shell such that the overall density 
of the tracer matches that of the glass particles.

\begin{figure}[htbp]
\includegraphics[width=\linewidth]{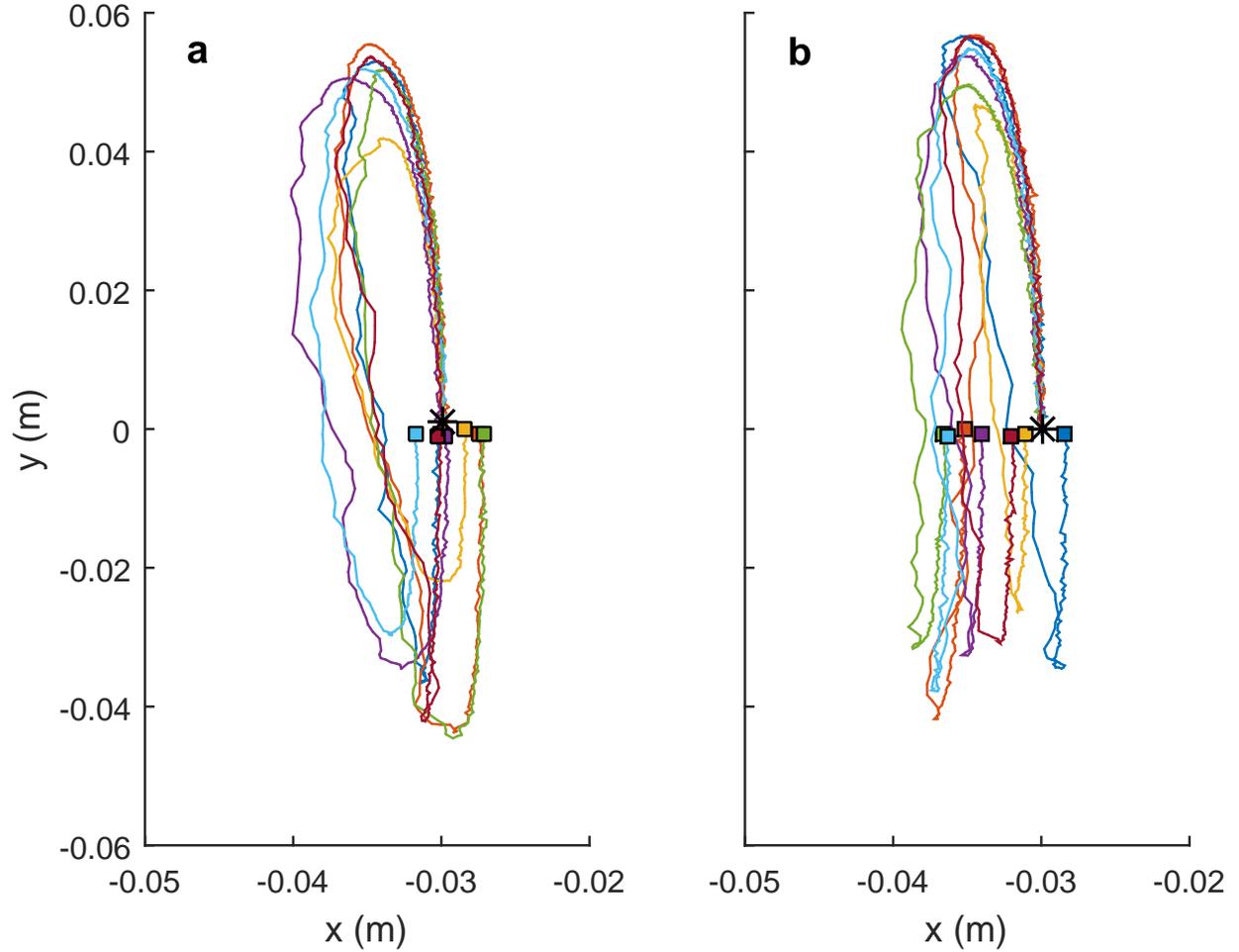}
\caption{(Color online) Individual trajectory segments (top view) from experiment of a 3~mm
tracer particle in a bed of 2~mm particles in a (a) smooth wall and in a (b) 
rough (2~mm particles) wall spherical tumbler filled at 30\%. 
Each trajectory segment
consists of one circulation through the fixed bed and the flowing layer with
a starting positions of $-3.5$~cm $\le x\le -2.5$~cm. The starting position (star)
of each trajectory was shifted to $x = −3$~ cm to make it easier to distinguish the
drift. The endpoint of each trajectory is marked with a square box. 
Note that the solid body portion of the trajectory appears curved due 
to the inherent depthwise geometric magnification in the X-ray image.}
\label{experiments}
\end{figure}

The experimental results are shown in Fig.~\ref{experiments} for 
7 randomly selected particle trajectories for each case, 
all starting with an axial position in the range $-3.5$~cm $\le x\le -2.5$~cm and directly below the axis
of rotation in the
fixed bed, but shifted axially in the figure to the same starting location.
The poleward drift in a smooth wall tumbler is close to zero in all cases 
(Fig.~\ref{experiments}(a), noting the trajectory endpoints marked with a 
square box), while it is up to 0.7~cm in the rough
wall tumbler (Fig.~\ref{experiments}(b)). While these results are limited 
in scope, they are consistent with the simulation results in 
Fig.~\ref{comparetrajroughsmooth} and our previous 
results for the amount of axial drift in a tumbler with smooth walls 
\cite{ZamanDOrtona13}. The curvature is less in both cases than occurs
in the simulations, perhaps due to different particles characteristics.
Nevertheless, these experiments clearly validate the simulation results 
in spite of substantial variation in the trajectories due to collisional
diffusion. That is, curvature of the trajectories, and hence displacement,
is larger for smooth tumbler walls, while axial drift is greater for rough
tumbler walls.
Due to the difficulty in performing these experiments, we focus on simulation
results in the remainder of the paper. 
Note however that we performed DEM simulations to match the conditions in the
experiments (3~mm tracer). These simulation results are consistent with the 
experiments and with DEM results tracking a single 2~mm (instead of 
3~mm) tracer particle, thus demonstrating that using a 3~mm tracer with 
otherwise 2~mm particles is a valid experimental approach.

\begin{figure}[htbp]
\includegraphics[width=0.88\linewidth]{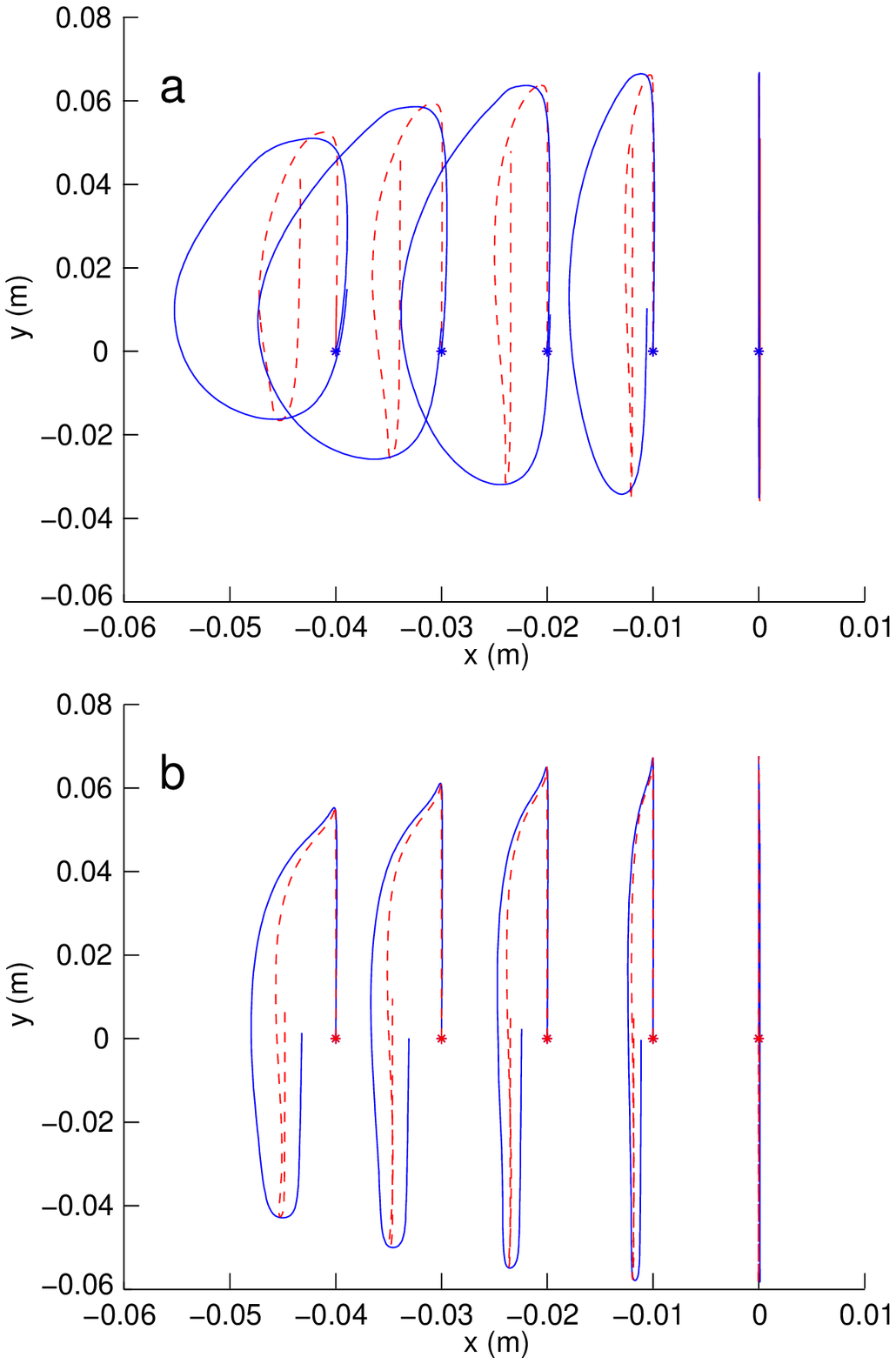}
\caption{(Color online) Comparison (top view) of the mean trajectories 
(initially 2~mm above the sphere wall) for 2~mm monodisperse
particles starting
from the same points for the smooth wall (blue continuous line) and the 2~mm 
(red dashed lines) rough wall a) in a tumbler filled at 20\%; b) in a 
tumbler filled at 50\%.}
\label{comparetraj2050ptop}
\end{figure}

Similar DEM simulation results occur for 2~mm particles in 20\% and 50\% full 
tumblers, as shown in Fig.~\ref{comparetraj2050ptop}, but the details are quite
different from the 30\% fill level results. In the smooth cases (blue curves), 
the curvature (displacement)
 of the trajectories
is larger 
for the 20\% volume fill and smaller for the 50\% volume fill
than the 30\% case, but the 
drift is largest for the 50\% case and near zero for a 20\% fill level. 
In the rough cases (red dashed curves) the displacement decreases slightly with
increasing fill level, and the drift increases slightly with fill level.
 As a result, the difference in the displacement and the 
drift between
smooth and rough walls is greatest for the low fill volumes and much
smaller for the 50\% fill volume.

We can now compare trajectories (Figs.~\ref{comparetrajroughsmooth} and 
\ref{comparetraj2050ptop}) 
with band deformations (Figs.~\ref{shear30} and \ref{shear50}) for 20\%, 30\% and 50\% fill
volumes. In the 20\% smooth case, there is almost no drift, and consequently, 
the band has little deformation, though it spreads due to diffusion.
 For other cases, the 
band deformation is directly linked with the drift of the trajectories: for 
any particular fill level, more drift results in greater band deformation for
the rough case.  The larger axial drift for rough walls is
a consequence of the particle trajectories curving toward the
pole in the upper part of the flowing layer, but not curving back toward
the equator in the lower part. In contrast, for smooth walls, the trajectories 
curve back toward the equator in the lower part of the flowing layer nearly as
much as they curved toward the pole in the upper part, particularly 
for lower fill levels.  

\begin{figure}[htbp]
\includegraphics[width=0.83\linewidth]{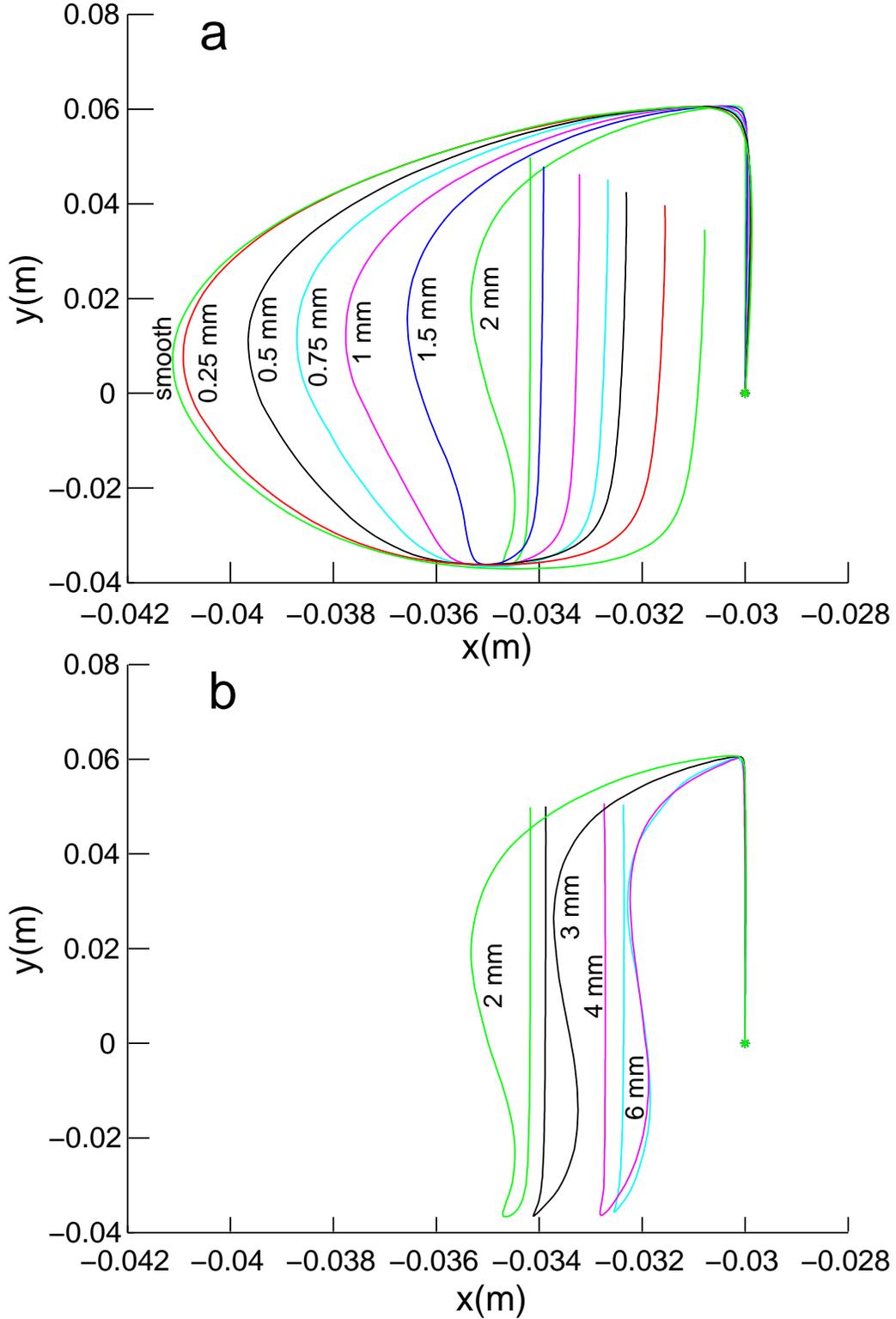}
\caption{(Color online) Comparison (top view) of the mean trajectories 
for particles starting from $x=-0.03$~m and 2~mm above the sphere wall 
for roughnesses ranging a) from a smooth wall
to 2~mm rough wall, b) from 2~mm to 6~mm rough walls. The horizontal axis is stretched 
compared to the vertical axis. The tumbler fill volume is 30\%.}
\label{compareroughtop0-6}
\end{figure}

The curvature of the trajectories for wall roughnesses 
ranging from smooth to 2~mm decreases monotonically, 
as shown in Fig.~\ref{compareroughtop0-6}(a). Thus, smooth
walls result in more curved trajectories with little drift, and rougher
walls result in less curved trajectories with more drift.
Larger roughnesses (Fig.~\ref{compareroughtop0-6}(b)) induce a modification in the flow
trajectories from simple curves to ones in which the 
curvature reverses, which is linked 
with the reduced slip of the flowing particles at the wall, as will be shown 
below. 
Furthermore, both
the displacement and drift decrease when the wall particle size
exceeds the flowing particle size of 2~mm.

\begin{figure}[htbp]
\includegraphics[width=0.95\linewidth]{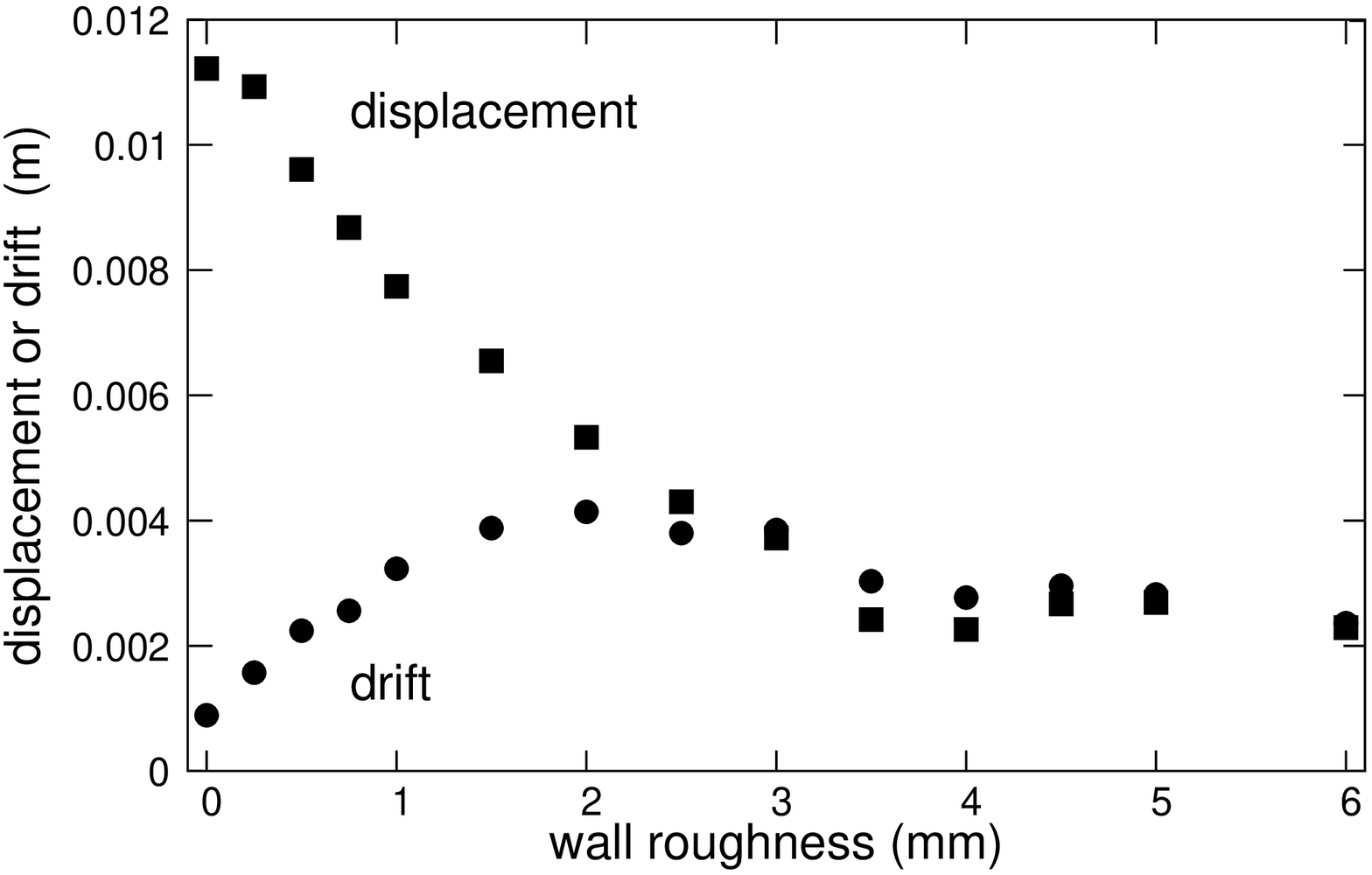}
\caption{Curvature (maximum displacement in one trajectory circulation) of 
the trajectories and axial drift 
for tumbler wall roughnesses ranging from smooth (0 mm) to 
6~mm rough walls for trajectories starting at $x=-0.03$~m in a 30\% full 
tumbler.} 
\label{roughness5}
\end{figure}

Drift becomes
progressively comparable to displacement as the roughness increases,
as shown in Fig.~\ref{roughness5} where drift and displacement 
are plotted versus wall roughness. 
The displacement decreases until a roughness around 4~mm, above which it
slightly increases
and then remains constant. This 
evolution is very similar to the case of a 
rough incline in that the maximum friction occurs for a roughness 
of the wall corresponding to wall particles approximately twice the size 
of the flowing particles, while for larger roughnesses, 
the friction slightly decreases and then reaches a constant value  \cite{GoujonThomas03}.

The dependence of the drift on the wall particle size 
is complex, since it results from both the
displacement toward the pole and the return curvature back toward the equator.
The drift increases with increasing roughness up to 2~mm roughness and
then decreases slightly so that it is nearly the same as the displacement. 
As noted earlier, the trajectories curve toward the pole in the
upper portion of the flowing layer for both smooth and rough walls, but
the trajectories do not curve back toward the equator in the lower portion
of the flowing layer for walls of 2~mm roughness and greater, as shown in 
Fig.~\ref{compareroughtop0-6}(b). Similar results occur for 20\% and 50\% fill
levels.

\subsection{Dependence on the depth in the flowing layer}

To fully understand the nature of the band deformations shown in 
Figs.~\ref{drift1mm}-\ref{shear50}, it is helpful to examine the 
trajectories at different depths in the flowing layer.
\begin{figure}[htbp]
\includegraphics[width=0.99\linewidth]{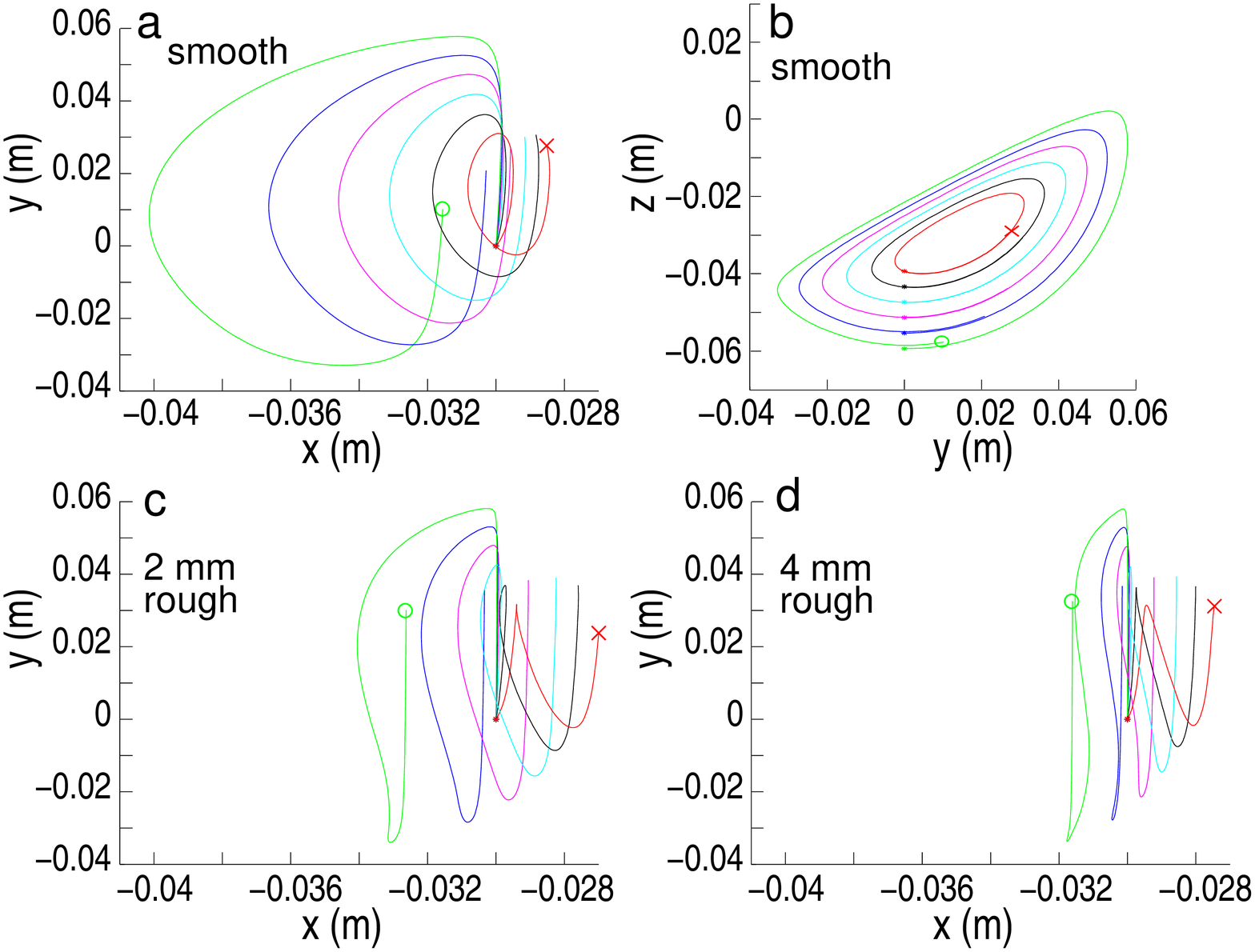}
\caption{(Color online) Mean trajectories in a spherical tumbler filled at 30\% with 2~mm particles flowing at different depth (starting 
positions 2, 6, 10, 14 and 18~mm above the sphere wall). (a) smooth wall sphere, top view; (b) smooth wall sphere
side view; (c) 2~mm rough wall, top view; (d) 4~mm rough wall, top view.
The curve ending with a circle is nearest the surface; the curve ending
with an X is deepest in the flowing layer.}
\label{depth30}
\end{figure}
Figure~\ref{depth30} shows the mean trajectories of particles starting from 
different vertical positions in the static bed for smooth and two
different rough walls.
In all cases, trajectories nearer the surface (e.g. green curve ending
with a circle) drift 
toward the
pole, while the deepest trajectories (red curve ending with an X) drift toward the equator.
For context, the trajectories viewed along the axis of rotation are shown 
in Fig.~\ref{depth30}(b) for smooth walls.
The drift toward the poles near the surface and toward the equator deep in
the flowing layer in Fig.~\ref{drift1mm} is consistent with previous studies 
\cite{ZamanDOrtona13} and explains the band deformation evident in Figs.~\ref{drift1mm}-\ref{shear50}.
The maximum
of drift difference (between the surface and the deepest trajectories) 
occurs for the 2~mm rough wall (Fig.~\ref{depth30}(c)) and is slightly less
for the 4~mm rough wall (Fig.~\ref{depth30}(d)). Indeed the largest deformation of the
band is obtained for wall roughnesses of 2~mm and greater
(Fig.~\ref{shear30}).

\begin{figure}[htbp]
\includegraphics[width=0.95\linewidth]{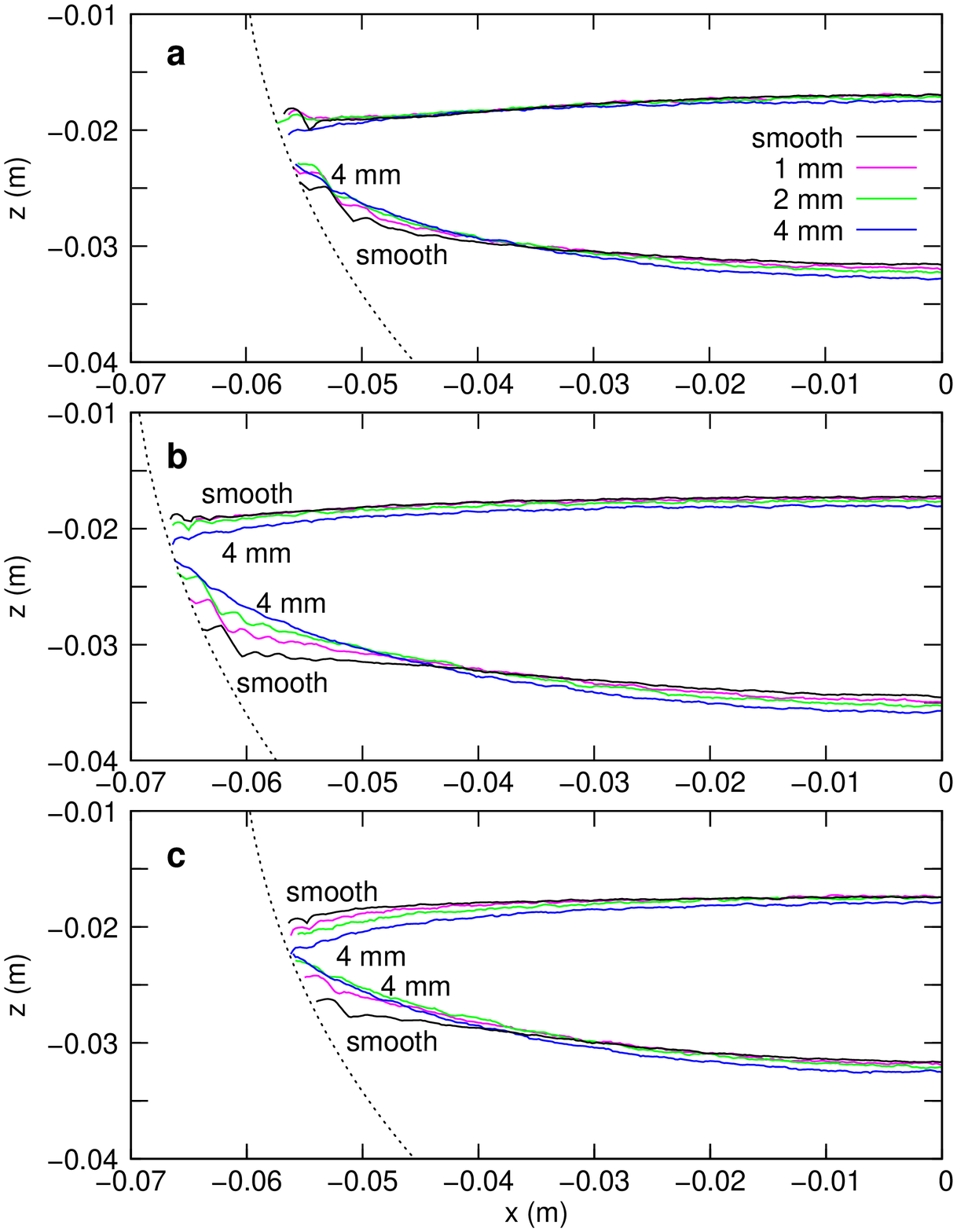}
\caption{(Color online) Topographies of the surface (top curves) and $v_y=0$ lower boundary
 (bottom
curves) of the flowing layer measured for three planes (a) $y=0.035$~m (top of 
the flow), (b) $y=0$ (middle of the flow), and (c) $y=-0.035$~m (bottom
of the flow). The tumbler is 
filled at 30\% with 2~mm particles for
smooth, 1~mm, 2~mm, and 4~mm rough walls. The dotted curves show the position
of the sphere wall.}
\label{vzero30p}
\end{figure}

Insight into the mechanism for the curved trajectories comes from close
examination of the flowing layer. Figure~\ref{vzero30p} shows the upper 
free surface and the lower boundary with the fixed bed of particles for
the flowing layer
at three vertical planes in the flow: upstream ($y=0.035$~m), 
middle corresponding to the axis of rotation ($y=0$), and
downstream ($y=-0.035$~m). The upper surface is based on the upper 
0.3 iso-compacity contour, while the lower surface is the position where
the out-of-plane velocity is zero: $v_y=0$. The jagged appearance for some
curves is a consequence of layers of particles adjacent to the walls.
 For the middle and downstream planes both the 
free surface 
and the boundary with the fixed bed become more curved near the wall (dotted
curve) as the
roughness of 
the wall increases. Consequently, for
larger roughnesses, the flowing layer thickness is reduced near the wall, while
for the smooth cases it is thicker indicating that particles may be
interacting with the wall differently.
In the upper part of the flow, $y=0.035$~m,
the flowing layer at the wall
has nearly the same thickness regardless of the wall roughness.
This behaviour is consistent with the trajectories of the particles
in the flow. 
In the upstream portion of the flowing layer, particles fall away from the 
wall, so the wall roughness has little impact.
In the downstream half of the flowing layer, particles move toward the 
curved wall and are thus sensitive to its roughness. 
 In the middle portion, particles flow parallel to the wall, and 
the behavior is intermediate between the upstream and downstream situations.

The dependence of the flowing layer thickness on the wall roughness provides 
insight into
the trajectories of particles.
Particles can easily move along a smooth wall, allowing a larger
displacement of the trajectories toward the pole. However, 
the flow is forced away from a rough wall toward the central zone of the tumbler, inducing 
trajectories with a smaller displacement toward the pole. For even larger
roughness, the flow at the wall is so small that the free surface in 
Fig.~\ref{vzero30p} curves downward, further reducing displacement toward 
the poles.  The impact of roughness is greatest in the downstream portion of 
the flowing layer as particles directly impact the wall, inducing 
a modification to trajectories in which the curvature 
reverses (Fig.~\ref{compareroughtop0-6}(b)).
In this way, 
the flow structure across the entire width of the flowing layer 
is modified by the roughness of the wall at its periphery.

\begin{figure}[htbp]
\includegraphics[width=\linewidth]{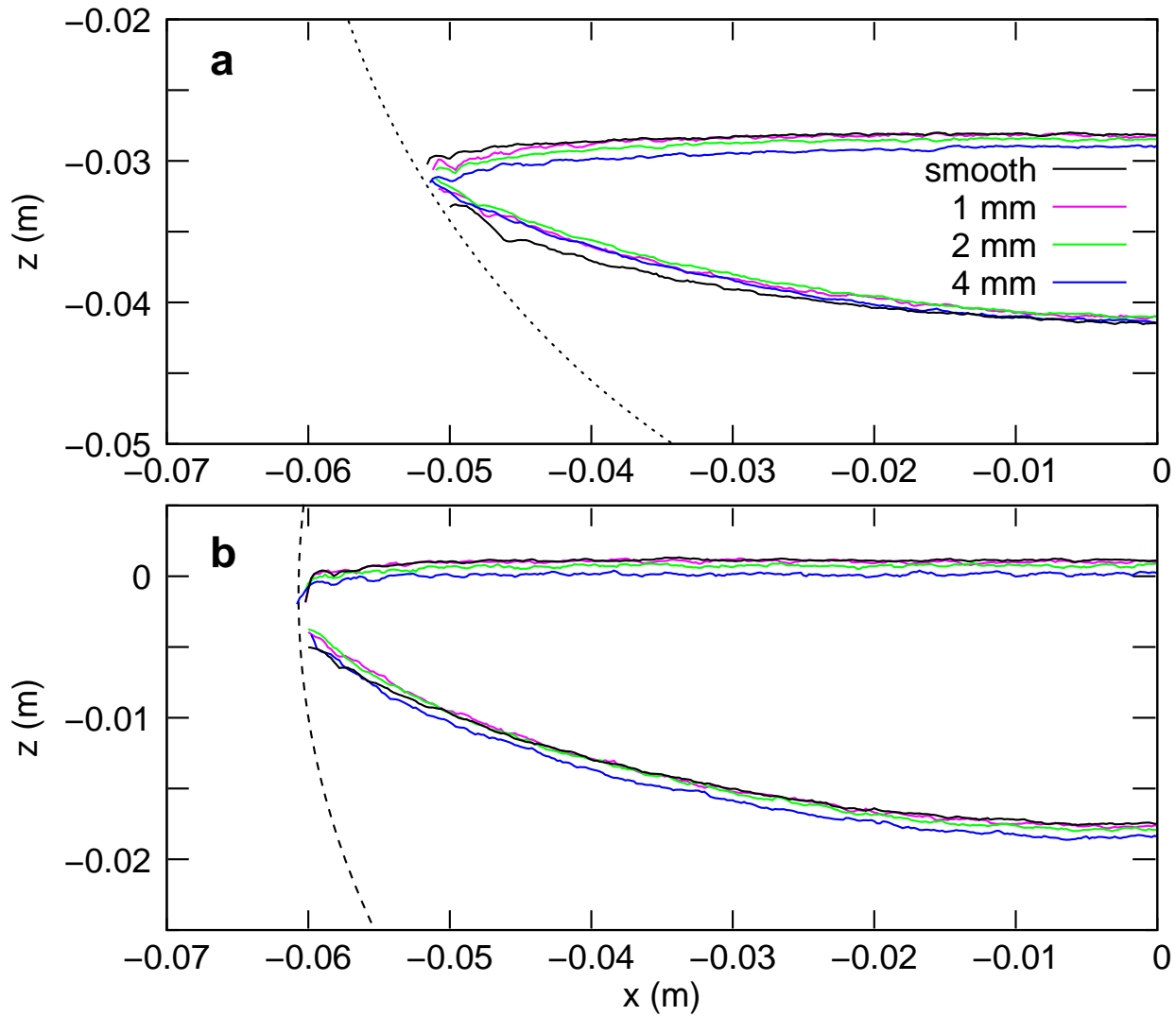}
\caption{(Color online) Topographies of the surface (top curves) and $v_y=0$ boundary (bottom
curves) of the granular flow measured in the plane $y=-0.035~m$ (bottom of
the flow). The tumbler is filled at (a) 20\% or (b) 50\% with 2~mm particles 
for smooth, 1~mm, 2~mm, and 4~mm rough walls. The dotted curve shows 
the position of the sphere wall.}
\label{vzero2050}
\end{figure}

Similar results occur for fill levels of 50\% and 20\% at the downstream plane
of the flow, but with slight differences.
For the 20\% case (Fig.~\ref{vzero2050}(a)) 
the free surface and boundary with the fixed bed are even more 
important than in the 30\% case. Like in the 30\% case, the flowing layer
thickness at the wall is greatly reduced, especially for larger wall roughness.
For the 50\% case (Fig.~\ref{vzero2050}(b)), the flowing layer thickness
very near the wall is nearly independent of wall roughness. Thus, wall
roughness has a much smaller effect on particle trajectories for a 50\%
fill level (see
Fig.~\ref{comparetraj2050ptop}(b)) than for lower fill levels.

\begin{figure}[htbp]
\includegraphics[width=0.9\linewidth]{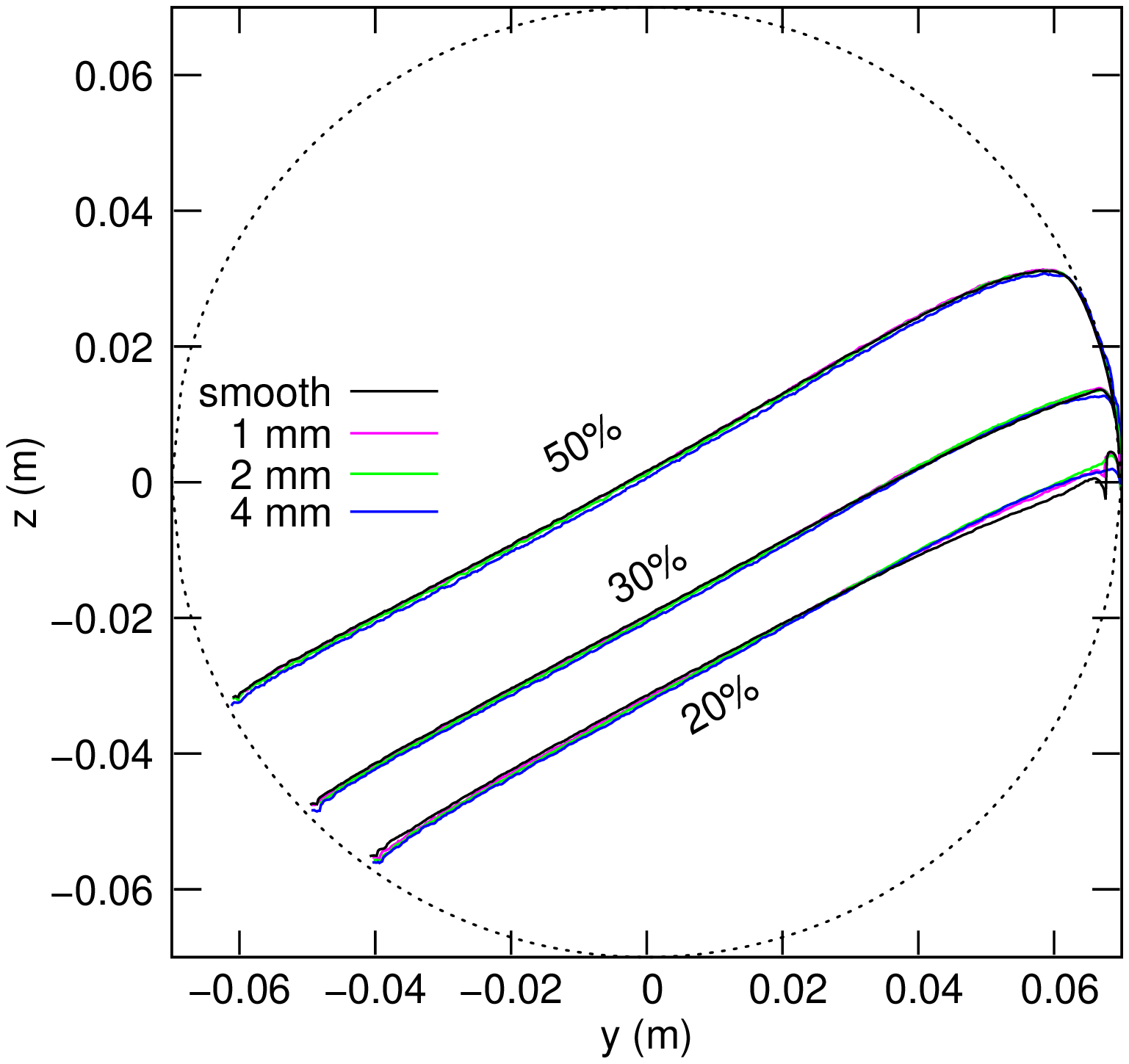}
\caption{(Color online) Profile of the free surface at the equator for monodisperse 2~mm
particles rotated at 15~rpm at $x=0$ for smooth,
1, 2, and 4~mm rough walls. Fill levels are 50\% (top curves), 30\%, 
and 20\% (bottom curves).
For each fill level, the four curves nearly overlay one another.}
\label{angle}
\end{figure}

For completeness, we note that
the roughness has no effect on the angle of repose.
Figure~\ref{angle} shows the surface profile at the equator ($x=0$)
for various fill levels and wall roughnesses. For each fill level, the
free surface profiles for different values of wall roughness nearly overlay 
each other. A small difference is
evident  in the upper portion of the 20\% fill case. In fact, at that location, the wall
is approximately vertical, allowing a monolayer of particles to form in contact
with the wall. This monolayer, which has been noted previously at low fill
levels in spherical tumblers \cite{ChenLueptow09}, reduces the height of 
the bed of particles just below it. It only occurs for the smooth wall and 
the 1~mm rough wall. 

\subsection{Surface velocity, slip at the wall}

\begin{figure}[htbp]
\includegraphics[width=0.90\linewidth]{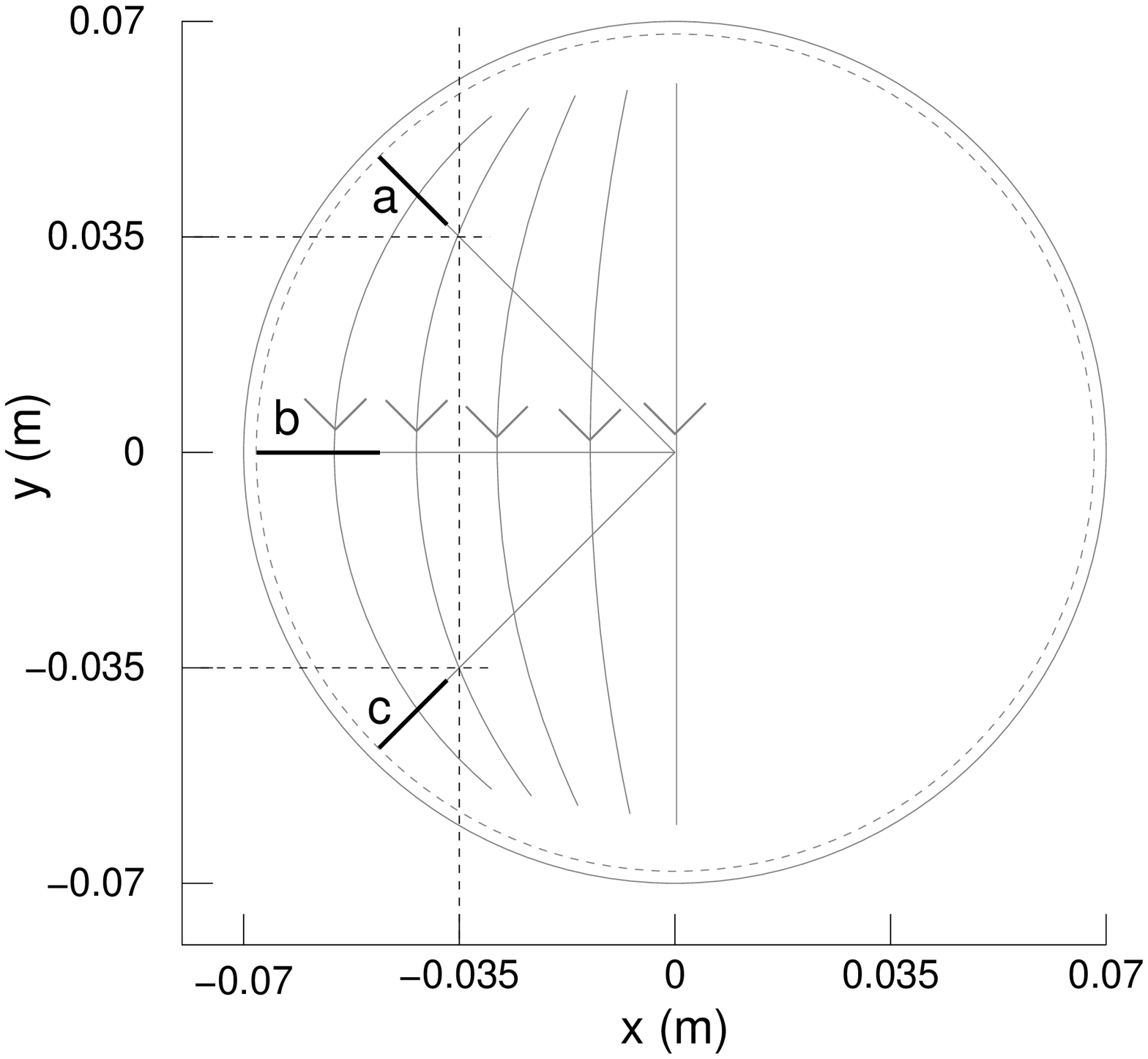}
\caption{Diagram showing radial lines along which the velocity profiles are 
measured.}
\label{diagparois}
\end{figure}

\begin{figure}[htbp]
\includegraphics[width=0.8\linewidth]{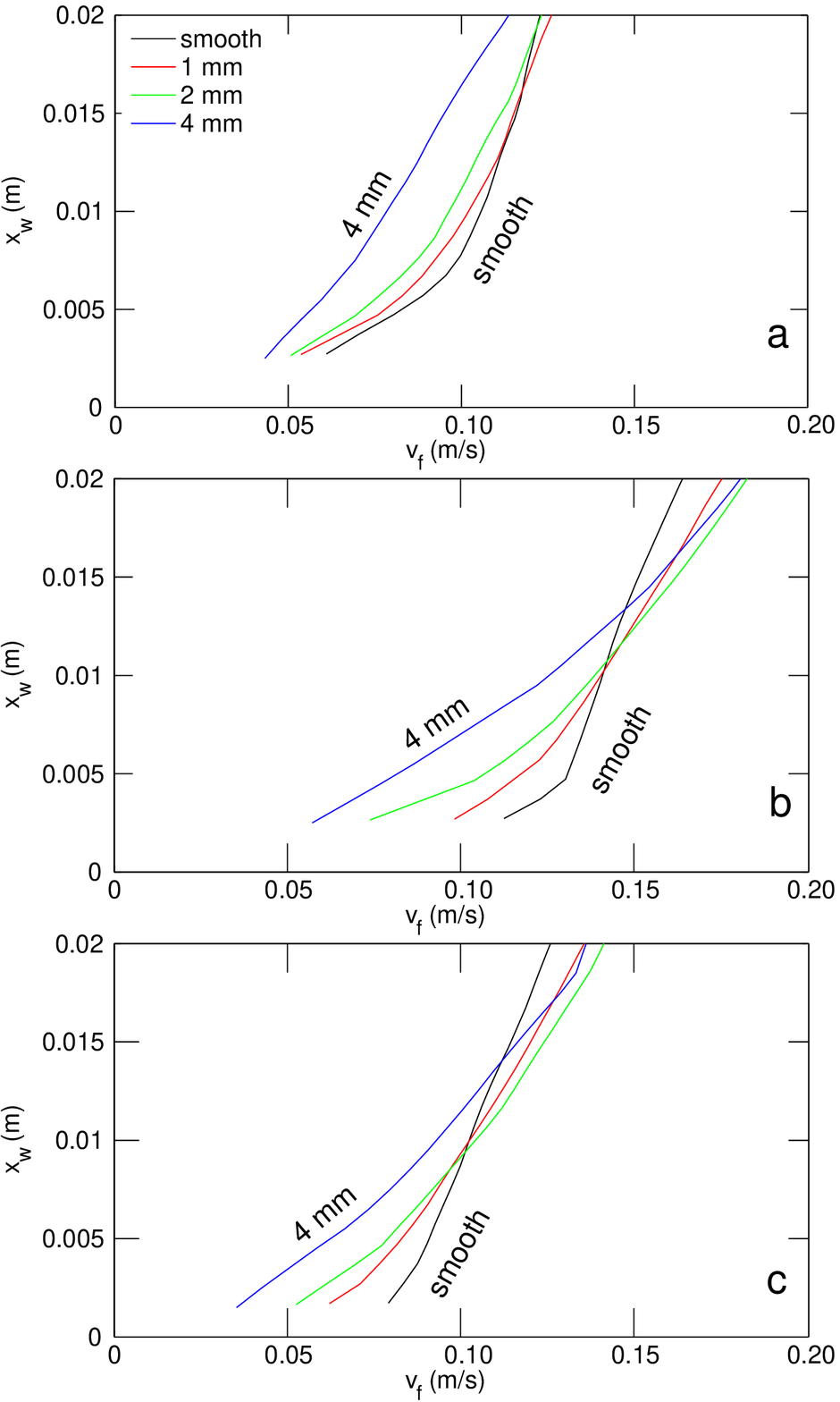}
\caption{(Color online) Velocity profiles for different wall roughnesses for a 30\% 
filled spherical tumbler at lines a, b, and c in the flowing layer.}
\label{wall30combine}
\end{figure}

Based on the dependence of the flowing layer thickness at the wall on wall 
roughness, it seems that the slip at the wall plays a major role in modifying
particle trajectories, changing the velocity profiles,
 and locally altering the thickness of the flowing layer. 
To better quantify the slip, we have determined the 
velocity near the wall along a radial coordinate $x_w$ extending normal from
the wall at
3 different radial lines a, b, and c (Fig.~\ref{diagparois}). 
The velocities $v_f$ are measured in the plane parallel to the 
free surface, but 2~mm below it.
Only the projection of the velocity perpendicular to the radial coordinate and 
in the plane is considered. The velocities plotted
in Fig.~\ref{wall30combine} are obtained using an 
interpolation (cubic spline) from the velocity field measured from
the simulation. As this velocity field is measured on a cubic grid, only 
cubes that are completely inside the sphere are used. Hence the
velocity profiles do not extend all the way to the wall. 

In the upper part of the flowing layer at line a (Fig.~\ref{wall30combine}(a)), the velocity very near the wall is similar for all roughnesses,
though it appears that slip at the wall is most likely for the smooth wall.
This is probably because 
the particles are moving away from the wall and thus less sensitive to wall 
roughness.
Figure~\ref{wall30combine}(b) shows the velocity profile 
at line b (mid-length of the flowing layer) where the flow is essentially parallel to the wall. Here the velocity near
the wall depends more strongly on the wall roughness with less roughness
corresponding to greater likelihood of slip at the wall.
Figure~\ref{wall30combine}(c) shows the velocity profile measured close to the
wall at line c. Again, wall roughness has significant impact on the 
velocity near the wall with any tendency for slip decreasing with roughness.
Similar results occur for other fill levels, though the 20\% fill level is  
slightly more sensitive to roughness and the 50\% fill level is slightly 
less sensitive to roughness in the downstream portion of the flowing layer.

\begin{figure}[htbp]
\includegraphics[width=0.9\linewidth]{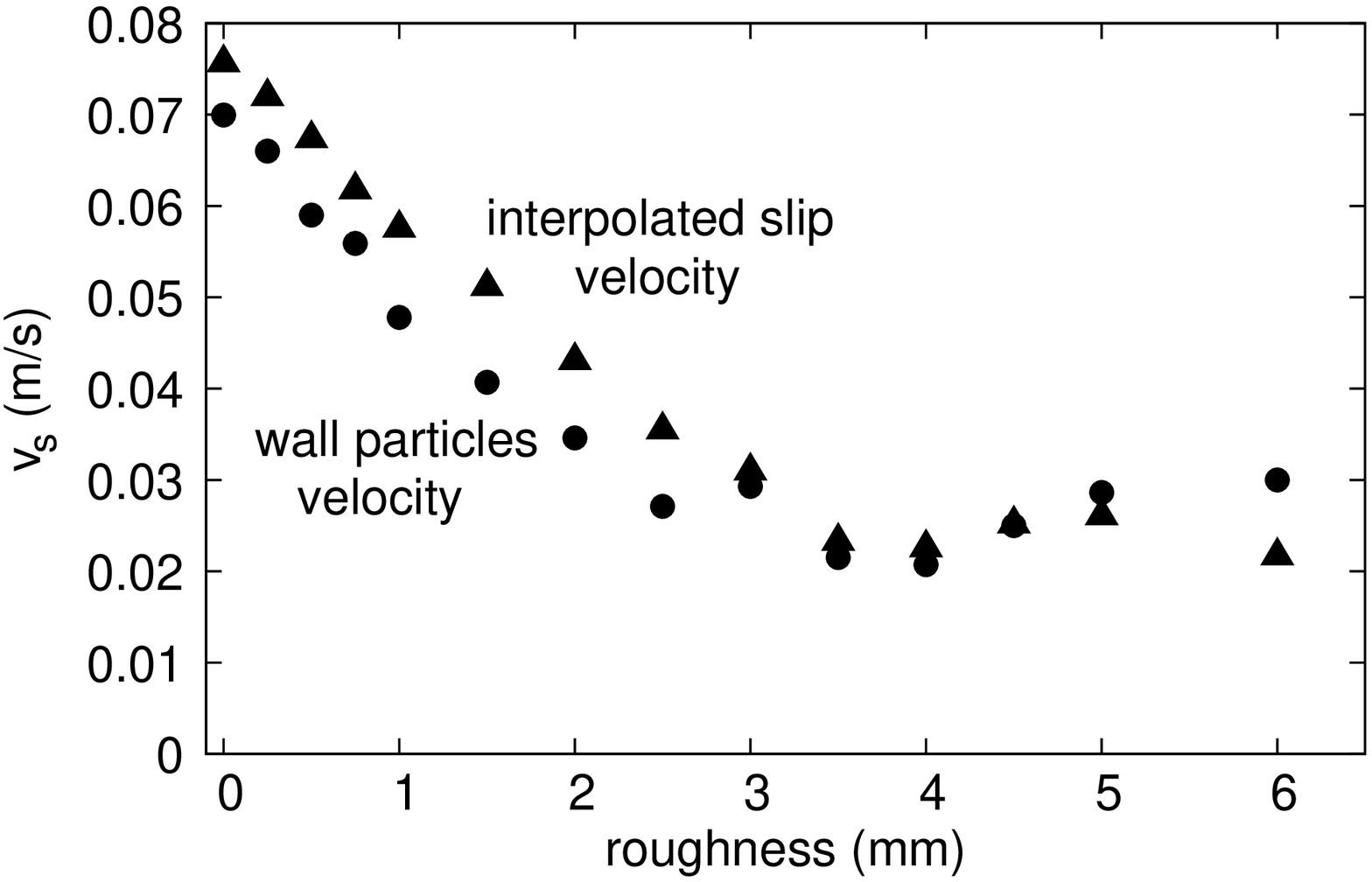}
\caption{Slip velocity based on extrapolating the velocity profiles
and by measuring the velocity of particles near the wall at
line c as a function of wall roughness for 30\% fill.}
\label{wallslip30}
\end{figure}

To quantify more precisely the influence of wall roughness on slip, we have
estimated the slip velocity at the wall $v_s$ in the downstream portion of the flowing layer at
c using two different methods.
First, the velocity profiles in Fig.~\ref{wall30combine} were extrapolated
with a line to the wall.  Second, the velocities of particles 
within a distance of 1.5 particle radii from the wall and close
to the free surface were directly measured. Regardless of the method
used to estimate the slip velocity, Fig.~\ref{wallslip30} shows that 
the slip at the wall decreases with roughness reaching a limit value
at around 3~mm wall roughness, above which increasing roughness does not further
 affect the wall slip. Note that this value is similar to the roughness
in Fig.~\ref{roughness5} where the displacement and drift become similar.
This result is likely analogous to the situation of an incline
in which the maximum friction inducing a minimum flow
velocity occurs when the roughness of the incline is approximatively twice the 
size of the flowing beads \cite{GoujonThomas03}.

\begin{figure}[htbp]
\includegraphics[width=0.9\linewidth]{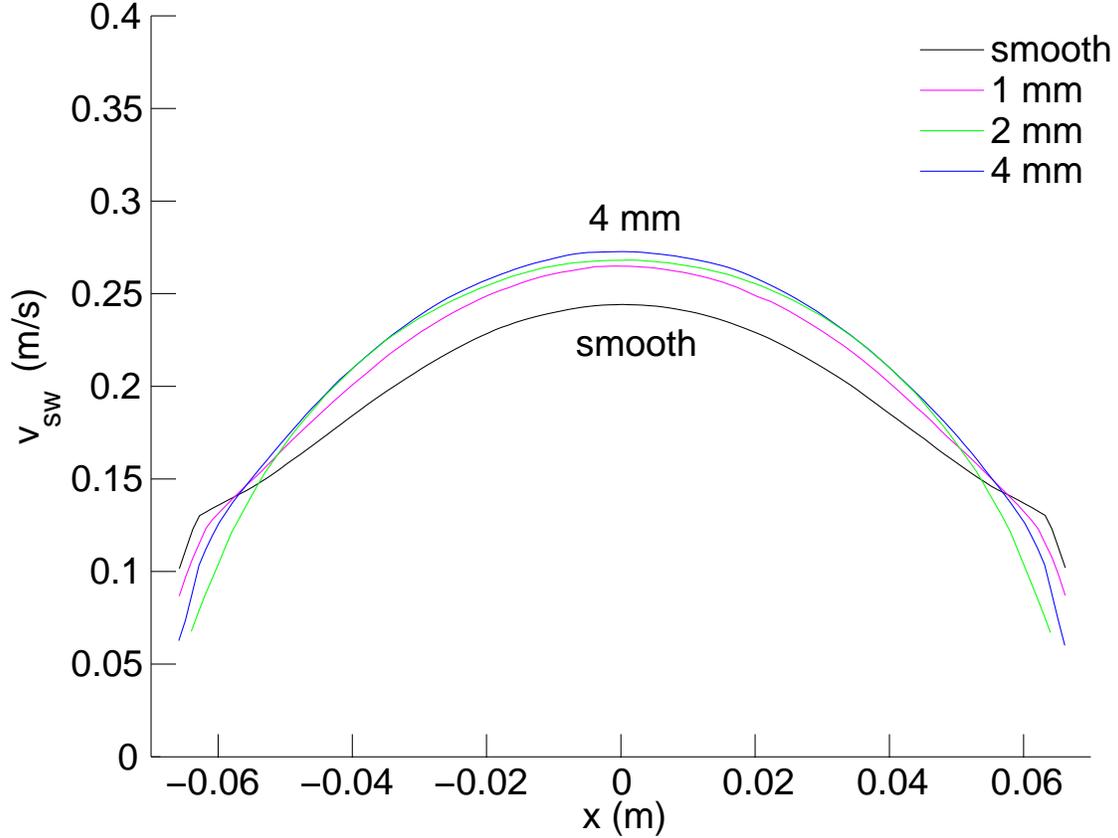}
\caption{(Color online) Streamwise velocity profiles measured 2~mm below the free surface 
at the $y=0$ plane for 4 different
roughnesses in a 30\% filled spherical tumbler.}
\label{meanprofile}
\end{figure}
 
The impact of the wall boundary condition on the overall flow is also evident
in the streamwise velocity (normal to the axis of rotation and in
the flowing layer surface) profile at the mid-length of the flowing layer
measured 2~mm below the surface (Fig.~\ref{meanprofile}). In the smooth
wall case, the velocity near the wall is higher due to slip. As a result, the 
velocity is lower
at the equator. Similar results occur deeper in the flowing
layer, as well. It is this difference in the streamwise velocity profiles, 
combined with the higher flux of particles at the equator as the flowing
layer thins for rough walls near its boundary (Figs.~\ref{vzero30p} and 
\ref{vzero2050}) that causes the longer recirculation times in the smooth 
case than in the rough case, noted with respect to Fig.~\ref{drift1mm}.

\subsection{Tumbler rotational speed and size ratio}

\begin{figure}[htbp]
\includegraphics[width=0.7\linewidth]{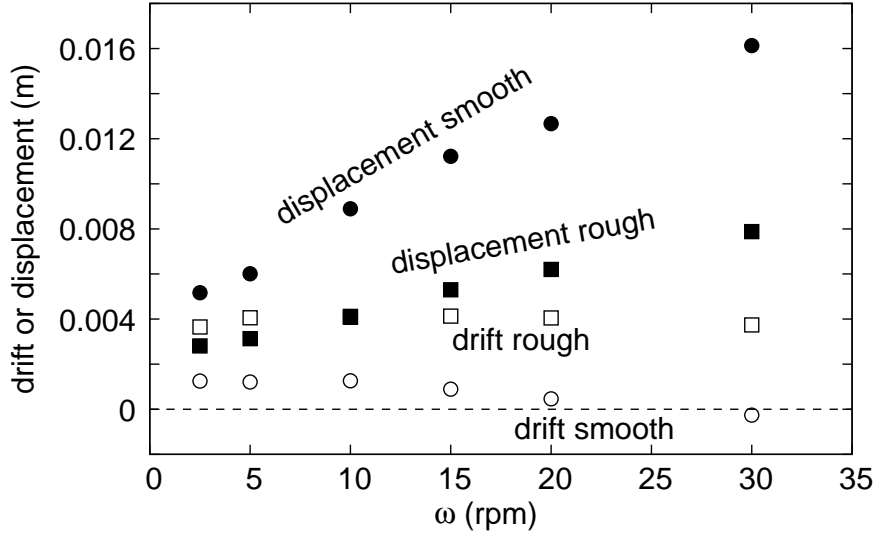}
\caption{Displacement and drift for trajectories starting at $x=-0.03$~m, $y=0$ and 2~mm above the bottom wall, for a spherical tumbler filled at 30\% with
2~mm particles at various rotation speeds.
 The walls are perfectly smooth
or 2~mm rough.} \label{velocitydrift}
\end{figure}

Finally, we vary both the rotational speed of the tumbler $\omega$  and the 
ratio of tumbler diameter to particle size ($D/d$) and find that drift and 
displacement of the trajectories occur even when these key parameters are 
varied. Consider first results 
for variations of the rotational speed, shown in Fig.~\ref{velocitydrift},
for 2~mm particles 
with smooth and 2~mm rough tumbler walls. 
The displacements for both smooth and rough tumbler walls increase 
with rotational speed, though the displacements for smooth walls are 
about twice that for rough tumbler walls. The drift varies little with 
rotational speed for rough tumbler walls, but decreases slightly with 
increasing rotational speed for smooth tumbler walls so that it becomes  
slightly negative by 30~rpm. The difference between smooth walls and rough
walls increases with rotation speed for both the drift and the displacement.

\begin{figure}[htbp]
\includegraphics[width=0.8\linewidth]{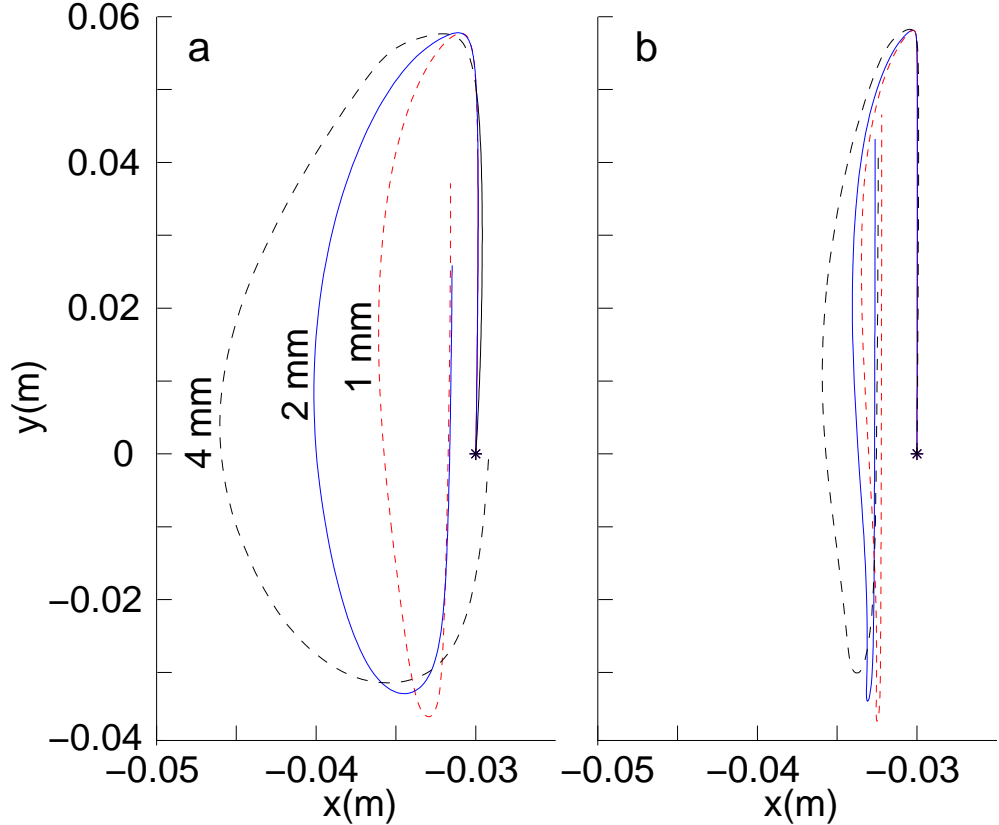}
\caption{(Color online) Mean trajectories (initially 4~mm above 
the sphere wall) in a spherical tumbler filled at 30\% with
particles 1~mm (red dashed), 2~mm (blue solid) and 4~mm (black dashed). The 
walls are (a) perfectly smooth or (b) rough made of
particles of the same size as the flowing particles.}
\label{wallsize}
\end{figure}

Perhaps more interesting is the effect of the tumbler diameter to particle 
size radio ($D/d$), given that wall roughness affects the flow surprisingly far 
from the wall for 2~mm particles in a 14~cm tumbler ($D/d=70$). To examine 
this, we plot particle trajectories for particles that are double and half 
the particle size used elsewhere in this paper ($d=4$~mm and $d=1$~mm) in 
Fig~\ref{wallsize}.
For both smooth
and rough walls, large particles have larger displacement than small 
particles, but the drift only weakly depends on particle size, except
for the largest particles.
Clearly, the wall roughness significantly affects the flow in all 
cases, though the impact decreases as the particle size relative to the 
tumbler size decreases.
\begin{figure}[htbp]
\includegraphics[width=0.8\linewidth]{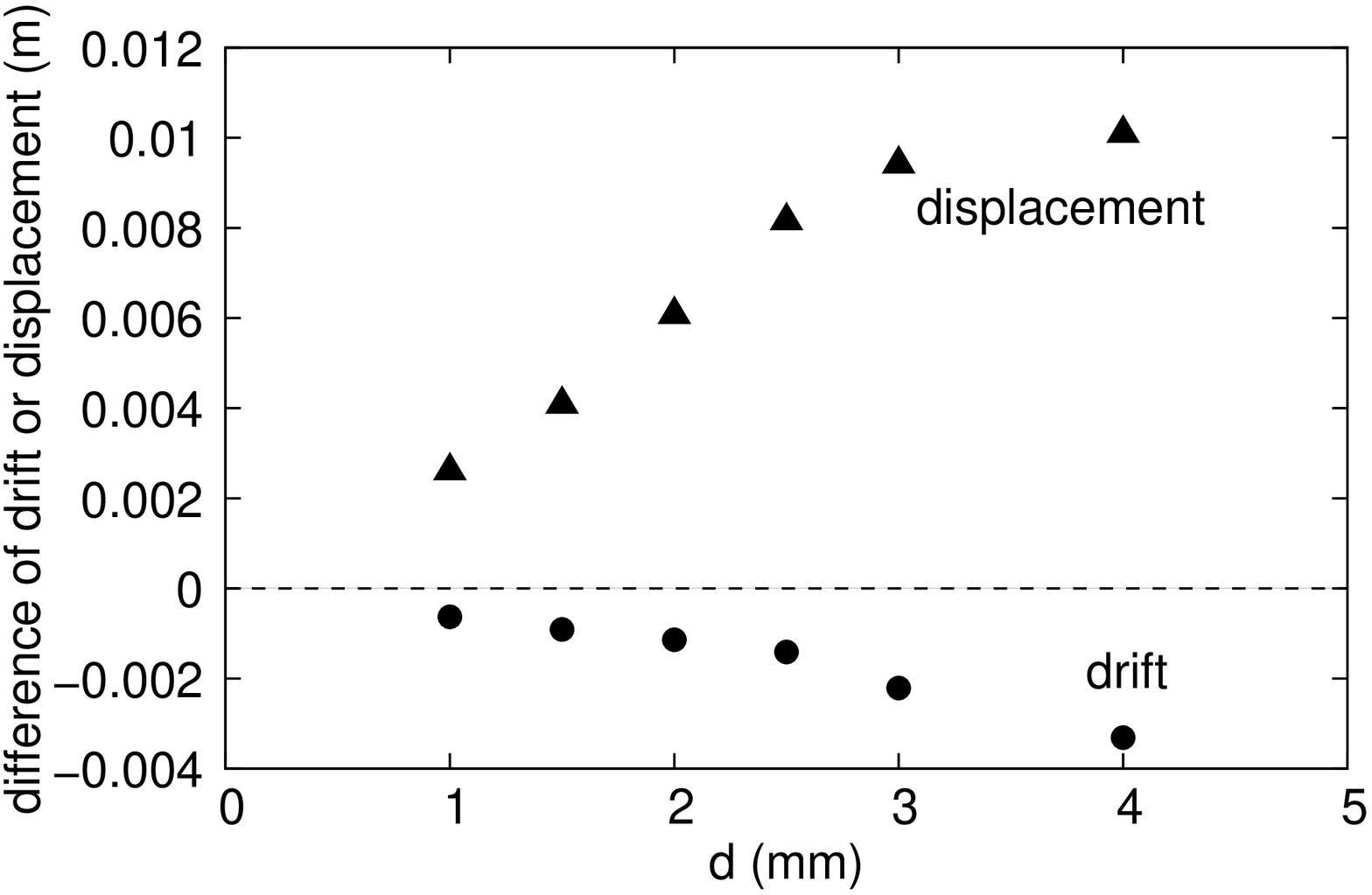}
\caption{Differences (smooth case minus rough case) for drift and 
displacement versus the size of the particles for trajectories
starting at $x=-0.03$~m, $y=0$, and 4~mm above the sphere wall for
a tumbler filled at 30\%.}
\label{sizedrift}
\end{figure}

Figure~\ref{sizedrift} shows the difference between smooth and rough wall cases for the
displacement and the drift for particle sizes ranging from 1~mm to 4~mm in a
14~cm tumbler. Both displacement and drift differences tend to
zero with decreasing particle size, but the decrease is nearly linear.
This suggests a remarkable persistence while
approaching zero, suggesting that the impact of wall roughness on flow
far from the wall persists even to fairly large values of $D/d$. Unfortunately, it
is quite difficult to confirm this, since simulations for particles smaller 
than 1~mm require a very large number of particles and very long computation
times.

\section{Conclusions}

The impact of wall roughness on details of the flow, even far from the bounding
wall, can be significant.  In a spherical tumbler, wall roughness combined 
with a curvature
of the wall determine the degree of slip at the wall, which in turn affects the
flowing layer thickness near the wall.  With rough walls and wall orientation
such that particles impinge on the wall (the downstream portion of the flowing
layer in the spherical tumbler), the resulting decrease in the flowing layer 
thickness near the wall forces the downstream flux of particles away from the
wall. This alters the particle trajectories and the flow even at the equator of
the tumbler.  On the other hand, smooth walls and wall orientation where 
particles fall away from the wall result in a higher slip velocity at the wall 
and consequently, less impact on the flowing layer thickness near the wall. 
 This allows a higher flux of particles near the wall.

The implications for DEM simulations are significant.  Even in situations 
where the
walls contact only the periphery of the flow, such as the case of a spherical
tumbler, the choice of wall roughness is 
critical. Preliminary results suggest a similar situation for
cylindrical tumblers. Of course, the wall roughness also affects other
 granular flows, such as straight chute flows,
but in these cases the lower wall directly contacts the bottom of the
flowing layer, so its impact is not surprising. 
The global impact of the peripheral flow boundaries may also play a role in
situations in which segregation occurs.  For instance, 
bidisperse segregation in spherical tumblers \cite{ChenLueptow09,NajiStannarius09} and even band formation 
in bidisperse flows in cylindrical tumblers, where the initial band formation 
seems to be driven by friction at the cylinder endwalls 
\cite{ChenLueptow11,ChenLueptow10}, may be 
strongly affected by the wall roughness far from the bounding walls.  
We are investigating these issues further.

\section*{Acknowledgments}

The effort of Z.Z. on this project was funded by NSF Grant No. CMMI-1435065.\\

\end{document}